\documentclass[12pt,preprint]{emulateapj}
\usepackage{epsf}
\usepackage{epsfig}
\usepackage{psfig}

\newcommand{\ltsima}{$\buildrel<\over\sim$}
\newcommand{\lapprox}{\lower.5ex\hbox{\ltsima}}
\newcommand{\msun}{M$_{\odot}$}
\newcommand{\kband}{{\it K$_s$}-band}

\shorttitle{The Mass Assembly History of Field Galaxies}
\shortauthors{Bundy et al.}

\begin{document}

\title{The Mass Assembly History of Field Galaxies: Detection of an Evolving Mass
Limit for Star Forming Galaxies}

\author{Kevin Bundy\altaffilmark{1}, Richard S. Ellis\altaffilmark{1},
  Christopher J. Conselice\altaffilmark{1}, James
  E. Taylor\altaffilmark{1}, Michael C. Cooper\altaffilmark{2},
  Christopher N. A. Willmer\altaffilmark{3}, Benjamin
  J. Weiner\altaffilmark{4}, Alison L. Coil\altaffilmark{2,5}, Kai
  G. Noeske\altaffilmark{6}, Peter R. M. Eisenhardt\altaffilmark{7}}

\email{kbundy@astro.caltech.edu, rse@atro.caltech.edu,
  cc@astro.caltech.edu, jet@astro.caltech.edu,
  cooper@astron.berkeley.edu, cnaw@as.arizona.edu, bjw@ucolick.org,
  acoil@as.arizona.edu, kai@ucolick.org, prme@kromos.jpl.nasa.gov}

\altaffiltext{1}{Caltech MC 105--24, 1201 E. California Blvd., Pasadena,  
CA 91125}
\altaffiltext{2}{Department of Astronomy, University of California at
   Berkeley, MC 3411, Berkeley, CA 94720}
\altaffiltext{3}{Steward Observatory, University of Arizona,
 Tucson, AZ 98721}
\altaffiltext{4}{Department of Astronomy, University of Maryland,
   College Park, MD 20742}
\altaffiltext{5}{Hubble Fellow, Steward Observatory, University of Arizona, Tucson, AZ 85721}
\altaffiltext{6}{UCO/Lick Observatory, University of California, Santa Cruz, CA 95064}
\altaffiltext{7}{Jet Propulsion Laboratory, California Institute of Technology, Pasadena, CA 91109}

\begin{abstract}

We characterize the mass-dependent evolution in a large sample of more
than 8,000 galaxies using spectroscopic redshifts drawn from the DEEP2
Galaxy Redshift Survey in the range $0.4 < z < 1.4$ and stellar masses
calculated from $K$-band photometry obtained at Palomar Observatory.
This sample spans more than 1.5 square degrees in four independent
fields.  Using restframe $(U-B)$ color and [OII] equivalent widths, we
distinguish star-forming from passive populations in order to explore
the nature of ``downsizing''---a pattern in which the sites of active
star formation shift from high mass galaxies at early times to lower
mass systems at later epochs.  Over the redshift range probed, we
identify a mass limit, $M_Q$, above which star formation appears to be
quenched. The physical mechanisms responsible for downsizing can thus be
empirically quantified by charting the evolution in this threshold mass.
We find that $M_Q$ decreases with time by a factor of $\approx$3 across
the redshift range sampled according to $M_Q \propto (1+z)^{3.5}$.  We
demonstrate that this behavior is quite robust to the use of various
indicators of star formation activity, including morphological type. To
further constrain possible quenching mechanisms, we investigate how this
downsizing signal depends on local galaxy environment using the
projected 3$^{rd}$-nearest-neighbor statistic $D_{p,3}$ which is
particularly well-suited for large spectroscopic samples. For the
majority of galaxies in regions near the median density, there is no
significant correlation between downsizing and environment.  However, a
trend is observed in the comparison between more extreme environments
that are more than 3 times overdense or underdense relative to the
median.  Here, we find that downsizing is accelerated in overdense
regions which host higher numbers of massive, early-type galaxies and
fewer late-types as compared to the underdense regions.  Our results
significantly constrain recent suggestions for the origin of downsizing
and indicate that the process for quenching star formation must,
primarily, be internally driven. By quantifying both the time and
density dependence of downsizing, our survey provides a valuable
benchmark for galaxy models incorporating baryon physics.

\end{abstract}


\section{Introduction}

The redshift interval from $z \approx 1$ to $z=0$ accounts for roughly
half of the age of the universe and provides a valuable baseline over
which to study the final stages of galaxy assembly. From many surveys
spanning this redshift range, it is now well-established that the global
star formation rate (SFR) declines by an order of magnitude
\citep[e.g.,][]{broadhurst92, lilly96, cowie99, flores99, wilson02}.  An
interesting characteristic of this evolution in SFR is the fact that
sites of active star formation shift from including high mass galaxies
at $z \gtrsim 1$ to only lower mass galaxies subsequently. This pattern
of star formation, referred to by \citet{cowie96} as ``downsizing,''
seems contrary to the precepts of hierarchical structure formation and
so understanding the physical processes that drive it is an important
problem in galaxy formation.

The observational evidence for the downsizing of star formation activity
is now quite extensive.  The primary evidence comes from field surveys
encompassing all classes of galaxies to $z \approx 1$ and beyond
\citep{BE00, bell05:SFR, bauer05, juneau05, faber05, borch06}. However, the
trends are also seen in specific populations such as field spheroidals,
both in their stellar mass functions \citep{fontana04,bundy05} and in
more precise fundamental plane studies \citep{treu05:downsize, vdwel05}
which track the evolving mass-to-light ratio as a function of dynamical
mass.  The latter studies find massive spheroidals completed the bulk of
their star formation to within a few percent prior to $z\simeq$1,
whereas lower mass ellipticals continue to grow by as much as 50\% in
terms of stellar mass after $z \sim 1$. Finally, detailed analyses of
the spectra of nearby galaxies suggest similar trends \citep{heavens04,
  jimenez05}.

Downsizing is important to understand as it signifies the role that
feedback plays in the mass-dependent evolution of galaxies.  As a
consequence, its physical origin has received much attention
theoretically.  Recent analytic work by \citet{dekel04}, for instance,
suggests that the distinction between star-forming and passive systems
can be understood via several characteristic mass thresholds governed by
the physics of clustering, shock heating and various feedback
processes. Some of these processes have been implemented in numerical
and semi-analytic models, including mass-dependent star formation rates
\citep{menci05}, regulation through feedback by supernovae
\citep[e.g.,][]{cole00, benson03, nagashima04} and active galactic
nuclei (AGN) \citep[e.g.,][]{silk98, granato04, dekel04, hopkins05:red,
  croton05, delucia05, scannapieco05, bower05, cattaneo06}.  However,
most models have, until now, primarily addressed the mass distinction
between star-forming and quiescent galaxies as defined at the present
epoch \citep[e.g.,][]{kauffmann03:bimodal}. Quantitative observational
measures of the {\em evolving mass dependence} via higher redshift data
have not been available.

This paper is concerned with undertaking a systematic study of how the
mass-dependent evolution of galaxies progresses over a wide range of
epochs.  The goal is to quantify the patterns by which evolution
proceeds as a basis for further modeling.  Does downsizing result
largely from the assembly history of massive early-type galaxies or is
there a decline in the fraction of massive star-forming systems?  In the
quenching of star formation, what are the primary processes responsible
and how are they related to the hierarchical framework of structure
assembly as envisioned in the CDM paradigm?  Does downsizing ultimately
result from internal physical processes localized within galaxies such
as star formation and AGN feedback, or is it caused by external effects
related to the immediate environment, such as ram pressure stripping and
encounters with nearby galaxies in groups and clusters?

In this paper, we combine the large spectroscopic sample contained in
the DEEP2 Galaxy Redshift Survey \citep{davis03} with stellar masses
based on extensive near-infrared imaging conducted at Palomar
Observatory to characterize the assembly history and evolution of
galaxies since $z \approx 1.2$.  Our primary goal is to quantify the
downsizing signal in physical terms and test its environmental
dependence so that it is possible to constrain the mechanisms
responsible.  A plan of the paper is as follows.  Section $\S$\ref{data}
presents the observations and characterizes the sample while
$\S$\ref{methods} describes our methods for measuring stellar masses,
star formation activity and environmental density.  We discuss how we
estimate errors in the derived mass functions in
$\S$\ref{mfn_description} and present our results in
$\S$\ref{results}. We discuss our interpretations of the results in
$\S$\ref{discussion} and conclude in $\S$\ref{conclusions}.  Where
necessary, we assume a Chabrier IMF \citep{chabrier03} and a standard
cosmological model with $\Omega_{\rm M}=0.3$, $\Omega_\Lambda=0.7$,
$H_0=100 h$ km~s$^{-1}$~Mpc$^{-1}$ and $h = 0.7$.

\section{Observations and Sample Description}\label{data}

Constraining the processes that govern downsizing requires a precise
measure of the evolving stellar mass function of galaxies
as a function of their physical state and environmental density.  Achieving
this ambitious goal requires multi-wavelength observations capable of
revealing quantities such as the ongoing star formation rate in a large
enough cosmic volume to reliably probe a range of environments.  Among
these requirements, two observational components are absolutely
essential: a large spectroscopic survey and near-IR photometry.

Spectroscopic redshifts not only precisely locate galaxies in space and time, but
enable the reliable determination of restframe quantities such as color
and luminosity.  This in turn allows for accurate comparisons to stellar
population templates which provide stellar mass-to-light ratios
($M_*/L$) and the opportunity to convert from luminosity to
stellar mass.  As demonstrated in \citet{bundy05}, relying on
photometric redshifts (photo-$z$) decreases the typical precision of
stellar mass estimates by more than a factor of three, with
occasional catastrophic failures that lead to errors as large as an 
order of magnitude.

Spectroscopic redshifts are also crucial for determining accurate
environmental densities \citep{cooper05}.  Even with the most optimistic
photometric redshift uncertainties of $\sigma_z = 0.02$---COMBO-17
specifies $\sigma_z \approx 0.03$ \citep{wolf03}---a comparison between
photo-$z$ density measurements and the real-space density in simulated
data sets gives a Spearman ranked correlation coefficient of only $\rho =
0.4$ \citep[where $\rho = 1$ signifies perfect correlation, see][]{cooper05}.
This uncertainty has the effect of smearing out the density signal in
all but the lowest density environments in photo-$z$ samples.  For the
spectroscopic DEEP2 sample used in this paper, $\sigma_z = 0.0001$ and
$\rho = 0.8$.

Spectroscopic observations also provide line diagnostics that can
discriminate the star formation activity occurring in galaxies.  Given
the various timescales involved, it is useful to identify ongoing or
recent star formation activity by considering various independent
methods including restframe $(U-B)$ color, the equivalent width of [OII]
$\lambda$3727, and galaxy morphology.  Comparisons between the different
indicators highlight specific differences between early- and late-type
populations defined in different ways.

In addition to spectroscopy, the second essential ingredient in this
paper is near-IR photometry.  As suggested by \citet{kauffmann98} and
first exploited by \citet{BE00}, near-infrared and especially $K$-band
photometry traces the bulk of the established stellar populations and
enables reliable stellar mass estimates for $z \lesssim 1.5$.  With the
addition of SED fitting from multi-band optical photometry, the
uncertainty in such estimates can be reduced to factors of 2--3 based on
$K$-band observations out to $z \approx 1.5$ \citep{BE00}.  The
importance of $K$-band observations is highlighted for galaxies with $z
> 0.75$, where the typical stellar mass uncertainty using the same
technique applied to photometry with the z-band as the reddest filter
would be worse by a factor of 3--4 (Bundy et al., in preparation).
Uncertainties from optical mass estimates become even less secure as the
redshift increases and increasingly bluer portions of the restframe SED
are shifted into the reddest filter bands.  Thus, the combined lack of
$K$-band photometry and spectroscopic redshifts leads to stellar mass
errors greater than factors of 5--10 with catastrophic failures off by
nearly two orders of magnitude.

Because of these factors, the combination of the DEEP2 Galaxy Redshift Survey
with panoramic IR imaging from Palomar Observatory represents the ideal
(and perhaps only) data set  for tracing the evolution of
accurately measured stellar masses and various indicators of star
formation activity across different environments to $z \approx 1.4$.  We 
provide details on the specific components of this data set below.

\subsection{DEEP2 Spectroscopy and Photometry}\label{DEEP2}

The DEEP2 Galaxy Redshift Survey \citep{davis03} utilizes the DEIMOS
spectrograph \citep{faber03} on the Keck-II telescope and aims to
measure $\sim$40,000 spectroscopic redshifts with $z \lesssim 1.5$ for
galaxies with $R_{AB} \leq 24.1$.  The survey samples four
widely-separated regions, covering a total area of 3.5 square degrees.
Targeting of the spectroscopic sample was based on $BRI$ photometry
obtained at the Canada--France--Hawaii Telescope (CFHT) with the
12K$\times$8K mosaic camera \citep{cuillandre01}.  Catalogs selected in
the $R$-band were constructed using the {\sc imcat} photometric package
\citep{kaiser95} and reach a limiting magnitude of $R_{AB} \sim 25.5$
\citep{coil04}.  The photometric calibration was computed with respect
to Sloan Digital Sky Survey (SDSS) observations which overlap a portion
of the CFHT sample.  The observed colors, used to estimate the inferred
restframe $(U-B)$ colors in this paper, were measured using apertures
defined by the object size in the $R$-band image.  Further details on
the CFHT photometry can be found in \citet{coil04}.

Sources were targeted for DEIMOS spectroscopy based on several criteria.
As determined by magnitude, color, and size, objects in the photometric
catalog were assigned a probability, $P_{gal}$, for being a galaxy.
Redshift targets were required to have $P_{gal} > 20\%$ with magnitudes
in the range, $18.5 \leq R_{AB} \leq 24.1$ \citep[see][]{willmer05}.  In
DEEP2 Fields 2--4, targets were also color-selected in $(B-R)$
vs. $(R-I)$ color space to lie predominantly at redshifts greater than
0.7.  Details on the color cuts will be presented in Faber et al. (in
preparation).  These selection criteria successfully recover 97\% of the
$R_{AB} \leq 24.1$ population at $z > 0.75$ with only $\sim$10\%
contamination from lower redshift galaxies \citep{davis05}.  The
redshift survey in these three fields is now complete, providing a total
of 21592 successful redshifts over 3 square degrees.  The last field,
the Extended Groth Strip (EGS) covers 0.5 square degrees and is
currently 75\% complete with a total of 9501 galaxies in the range $0 <
z < 1.5$.  The selection of targets in EGS is magnitude-limited,
providing a valuable sample for verifying the success of the color
selection used in Fields 2--4.  The sampling rate of galaxies satisfying
the target criteria is $\sim$60\% and DEEP2 galaxies from all 4 fields
are included in this paper.

DEEP2 redshifts were determined by comparing various templates to  
observed spectra as well as fitting specific spectral features.  This process is
interactive and will be described in Newman et al.~(in preparation).
Spectroscopic redshifts are used in this paper only when two or more
features have been identified in a given spectrum \citep[giving ``zquality''
  values $ \geq 3$,][]{faber05}.  The fraction of
objects for which this process fails to give a reliable redshift is
roughly 30\% and is dominated by faint blue galaxies, the majority of
which are beyond $z \sim 1.4$ where the [OII] $\lambda$3727 feature is
redshifted beyond the DEEP2 spectral wavelength range \citep{willmer05}.
More details on the redshift success rate are provided in
\citet{willmer05}, and \citet{coil04} further discuss the survey
strategy and spectroscopic observations.

\subsection{Palomar Near-IR Imaging}

Motivated largely by this study, we have conducted an extensive panoramic 
imaging survey of all four DEEP2 fields with the  Wide Field Infrared 
Camera \citep[WIRC,][]{wilson03} on the 5m Hale Telescope
at Palomar Observatory.  We describe the salient points of this survey here 
and will provide further observational details in a later article 
(Bundy et al., in preparation).  

The Palomar survey commenced in fall 2002 and was completed after
65 nights of observing in October 2005. Using contiguously spaced pointings
(each with a camera field-of-view of 8\farcm6$\times$8\farcm6) tiled in a
3$\times$5 pattern, we mapped the central third of Fields 2--4 to a
median 80\% completeness depth greater than $K_{AB}=21.5$, with 5
pointings deeper than $K_{AB}=22.5$.  The imaging in Fields 2--4
accounts for 0.9 square degrees or 55\% of the Palomar \kband~survey.

The remainder of the data was taken in the EGS where the \kband~data
covers 0.7 square degrees, but to various depths. The EGS was considered
the highest priority field in view of the many ancillary
observations---including HST, Spitzer, and X-ray imaging---obtained
there. As the orientation of the WIRC camera is fixed on the sky and the
EGS traces a $\sim$45 degree strip 16\arcmin~wide, to fully cover the
spectroscopic field in the E--W direction requires rows of three WIRC
pointings.  The N--S direction requires about 12 different positions,
so, in total, 35 WIRC pointings were used to map the EGS in the \kband.
The deepest observations were obtained along the center of the strip
where there is complete overlap between WIRC and the
spectroscopically-surveyed area.  In these regions, the typical depth is
greater than $K_{AB}=22.5$.  The rest of the southern half of the EGS
reaches $K_{AB}=22.3$, while that for the northern half is complete to
$K_{AB}=21.7$.  Additional Palomar $J$-band observations were obtained
for most of the central strip of the EGS and for Fields 3 and 4.  These
provide $J$-band photometry for roughly half of the \kband~sample and
are useful in improving photometric redshifts.

Individual \kband~exposures at a given pointing were taken with an
integration time of 2 minutes using coadditions of either 4
$\times$ 30 seconds in average temperature conditions, 3 frames $\times$
40 seconds in cooler conditions, or 6 frames $\times$ 20 seconds in warm
conditions.  The exposures were dithered at 27 positions in a
non-repeating, pseudo-random $\approx$7\arcsec pattern and combined into
54-minute mosaics using a double-pass reduction pipeline developed
specifically for WIRC.  At a given pointing, individual mosaics were
often obtained on different nights and so may vary in terms of seeing, sky 
background levels and transparency.  Most \kband~pointings consist of more than
two independently-combined mosaics with the deepest pointings comprising
as many as 6 independent mosaics.  Mosaic coaddition was performed 
using an algorithm that optimizes the depth of the final image by applying 
weights based on the seeing, background, and transparency of the subframes.  
The final seeing FWHM in the \kband~images ranges from 0\farcs8 to 1\farcs2, 
and is typically 1\farcs0.  Photometric calibration was carried out by
referencing standard stars during photometric conditions and astrometric
registration was performed with respect to DEEP2 astrometry
\citep[see][]{coil04} using bright stars from the CFHT $R$-selected
catalog.  More details on the survey strategy and image processing are
presented in Bundy et al.~(in preparation).

After masking out the low signal-to-noise perimeter of the final
\kband~images, we used SExtractor \citep{bertin96} to detect and measure
the \kband~sources.  As an estimate of the total \kband~magnitudes,
which are used to derive the luminosity and stellar mass of individual
galaxies (see $\S$\ref{masscode}), we use the {\sc MAG\_AUTO} output
from SExtractor.  We do not adjust this Kron-like magnitude to account
for missing light in extended sources, which introduces two slight
biases in our total magnitudes.  First, the average difference between
the {\sc MAG\_AUTO} values of a given galaxy and its corresponding
magnitude measured in a 4\arcsec\ diameter aperture is $\approx$0.03 mag
fainter at the highest redshifts compared to the lowest redshifts in the
sample.  This effect leads to a systematic underestimate of 0.01 dex in
stellar masses at high-$z$ compared to low-$z$.  In addition, {\sc
  MAG\_AUTO} systematically underestimates the total magnitude of blue
galaxies (defined by restframe $U-B$ color, see \ref{U-B}) in the lowest
redshift interval as compared to red galaxies.  The magnitude of this
effect, which likely results from the more extended nature of typical
blue galaxies, is also $\approx$0.03 mag but is not significant in the
middle or high redshift intervals.  Considering these systematic effects
contribute less than 0.01 dex to the final stellar mass estimates, we do
not explicitly correct for them.

In addition to total magnitudes, we also use SExtractor to measure
aperture photometry in diameters of 2\arcsec, 3\arcsec, 4\arcsec, and
5\arcsec.  To determine the corresponding magnitudes of \kband~sources
in the CFHT $BRI$ and Palomar $J$-band images, we applied the IDL
photometry procedure, APER, to these data, placing apertures with the
same set of diameters at sky positions determined by the
\kband~detections (all magnitudes were corrected to airmass = 1).  This
method was adopted with the aim of producing a $K$-selected
catalog---about 25\% of the \kband~sources do not have optical
counterparts in the CFHT {\sc imcat} catalog.  Magnitudes measured with
APER are consistent with SExtractor aperture magnitudes, and
experimentation with color-color diagrams and photometric redshifts
demonstrated that the 2\arcsec~diameter colors exhibited the least
scatter.  We therefore use the 2\arcsec~aperture colors for fitting
template spectral energy distributions (SEDs) to constrain stellar mass
estimates, as well as to estimate photo-$z$'s for sources beyond the
DEEP2 spectroscopic limit.

Photometric errors and the \kband~detection limit of each image were
estimated by randomly inserting fake sources of known magnitude into
each \kband~image and recovering them with the same detection parameters
used for real objects.  The inserted objects were given Gaussian
profiles with a FWHM of 1\farcs3 to approximate the shape of slightly
extended, distant galaxies.  We define the detection limit as the
magnitude at which more than 80\% of the simulated sources are detected
by SExtractor.  Robust photometric errors based on simulations involving
thousands of fake sources were also determined for the $BRI$ and
$J$-band data.  These errors are used to determine the uncertainty of
the stellar masses and in the determination of photometric redshifts,
where required.

\subsection{The Primary Sample}

Given the various ingredients necessary for the data set, it is
helpful to construct separate samples based on the differing
completeness limits for the $R$-limited spectroscopic sample
and the $K$-limited Palomar catalog.

We will define the primary sample as that comprising galaxies with DEEP2
spectroscopic redshifts that are also detected in the Palomar
\kband~imaging.  Redshift counterparts were found by cross-referencing
the \kband~catalog with the DEEP2 redshift catalog.  A conservative
tolerance of 1\farcs1 in the difference between the \kband~and DEEP2
positions was used to select matches, although the offset for most
sources was less than 0\farcs5.  The relatively low surface density of
both catalogs assures that the number of spurious associations is less
than 1--2 percent.  The fraction of DEEP2 redshifts detected in the
\kband~varies from $\approx$65\% for \kband~depths near $K_{AB}=21.7$ to
$\approx$90\% for $K_{AB}=22.7$.  After removing the redshift survey
boundaries to allow for unbiased density estimates (see
$\S$\ref{density}), Fields 2, 3, and 4 host 953, 1168, and 1704 sample
galaxies respectively, all with secure spectroscopic redshifts in the
range $0.75 < z < 1.4$.  The fourth field, the EGS, contains 4770
galaxies with redshifts in the range $0.2 < z < 1.4$.

In the analysis that follows, we divide this sample into three broad
redshift intervals.  The first, $0.4 < z < 0.7$, contains 943 galaxies,
drawn entirely from the EGS while the second ($0.75 < z < 1.0$, 2210
galaxies) and third ($1.0 < z < 1.4$, 1430 galaxies) draw from all four
fields (see Table \ref{sample_table}).  Although the entire
spectroscopic sample was selected to have $R_{AB} \leq 24.1$, the
limiting \kband~magnitude was chosen separately for each redshift
interval.  Because the Palomar \kband~survey covers different areas to
different depths, the area and volume of a given subsample depend on
the \kband~limit that is applied.  This is advantageous and allows us to
choose limits for each redshift bin that maintain adequate statistics
and stellar mass completeness while producing subsamples that probe
similar cosmological volumes---an important consideration for making
environmental comparisons in a consistent fashion.  In the three
redshift bins, we select galaxies with secured \kband~detections
brighter than 21.8, 22.0, and 22.2 (AB), respectively.  In the standard
cosmology we have adopted with $h=0.7$, the areas sampled with these
limits result in comoving volumes of 0.5, 1.4, and 2.3 in units of
$10^{6}$ Mpc$^3$.

\begin{deluxetable*}{lcccccc}
\tablecaption{Sample statistics}

\tablewidth{0pt}
\tablecolumns{6}
\tablehead{
\colhead{} & \colhead{Primary Spec-$z$, $R_{AB} \leq 24.1$} & \colhead{} & \multicolumn{4}{c}{Photo-$z$ Supplemented, $R_{AB} \leq 25.1$} \\
\cline{2-2} \cline{4-7} \\
\colhead{Sample} & \colhead{Number} & \colhead{} & \colhead{Number} &
\colhead{$f_{{\rm spec\mbox{-}}z}$} & 
\colhead{$f_{{\rm ANN}z}$} & \colhead{$f_{\rm BPZ}$} \\
\cline{1-7} \\
\multicolumn{7}{c}{EGS Field}
}

\startdata
Full sample      & 2669 & & 8540 & 36\% & 51\% & 13\% \\
$0.4 < z < 0.7$  & 943  & & 3026 & 36\% & 62\% & 2\% \\
$0.75 < z < 1.0$ & 1003 & & 2801 & 42\% & 46\% & 12\% \\
$1.0 < z < 1.4$  & 723  & & 2713 & 29\% & 43\% & 28\% \\
\cutinhead{DEEP2 Fields 2--4}
Full sample      & 1914 & & 10156 & 21\% & 68\% & 11\% \\
$0.4 < z < 0.7$  & ---  & & 4264  & 2\%  & 97\% & 1\% \\
$0.75 < z < 1.0$ & 1207 & & 2865  & 42\% & 48\% & 10\% \\
$1.0 < z < 1.4$  & 707  & & 3027  & 27\% & 47\% & 26\% \\

\enddata
\label{sample_table}
\tablecomments{The listed values reflect cropping the DEEP2 survey
  borders to allow for accurate environmental density measurements (see
  $\S$\ref{density}) and, for the three redshift intervals,
  \kband~magnitude cuts of 21.8, 22.0, and 22.2 (AB).}
\end{deluxetable*}

\subsection{The Photo-$z$ Supplemented Sample}

As mentioned previously, photometric redshifts are insufficient for
accurate density measurements \citep{cooper05}, do not offer [OII]-based
SFR estimates, and significantly decrease the precision of stellar mass
\citep{bundy05} and restframe color estimates.  However, photometric
redshifts do offer the opportunity to augment the primary sample and
extend it to fainter magnitudes, providing a way to test for the effects
of incompleteness in the primary spectroscopic sample because of 
the various magnitude limits and selection procedures.  With this goal 
in mind, we constructed a comparison sample using the optical+near-IR photometry
described above to estimate both photometric redshifts and stellar
masses.  We discuss the importance of this comparison in interpreting
our results in section $\S$\ref{completeness}.

Photometric redshifts were derived in two ways.  First, because the
DEEP2 multi-slit masks do not target every available galaxy, there is a
substantial number of objects that satisfy the photometric criterion of
$R_{AB} \leq 24.1$ without spectroscopic redshifts.  These are ideal for
neural network photo-$z$ estimates because of the availability of a
large spectroscopic training set.  For these galaxies, we use ANN$z$
\citep{collister04} to measure photo-$z$'s, training the network with
the EGS spectroscopic sample so that the ANN$z$ results cover the full
range, $0.2 < z < 1.4$, without being subject to biases from any color
selection.  Based on comparisons to the spectroscopic samples in Fields
2--4, the ANN$z$ results are in excellent agreement with $\Delta z /
(1+z) \approx 0.07$.

While the $R_{AB} \leq 24.1$ ANN$z$ photo-$z$ results present a complete
sample, they do not contain fainter galaxies because no spectroscopic
sample is available to train them.  We therefore define a second sample
of galaxies with $24.1 < R_{AB} \leq 25.1$ and 3$\sigma$ detections in
the $BIK$ bands.  For these we use the BPZ photo-$z$ package
\citep{benitez00} and 2\farcs0 diameter aperture photometry, including
$J$-band where available.  We first optimize BPZ by comparing to
spectroscopic samples.  These tests reveal a systematic offset in the
BPZ results that likely arises from the assumed HDF-North prior
\citep{benitez00}.  We fit a linear relation to this offset and remove
it from all subsequent BPZ estimates.  This improves the spectroscopic
comparison to $\Delta z / (1+z) \approx 0.17$.  For sources with $R_{AB}
\leq 24.1$, we find good agreement between the ANN$z$ and BPZ results,
although the BPZ estimates exhibit a larger degree of scatter.

Using the ANN$z$ and BPZ results, we construct photo-$z$ supplemented
samples in each redshift bin with an $R$-band limit of $R_{AB} \leq
25.1$.  The first bin, with $0.4 < z < 0.7$, contains 7290 galaxies;
16\% have spectroscopic redshifts (available in the EGS only), 83\% have
ANN$z$ photo-$z$'s, and only 1\% have BPZ photo-$z$'s.  The second bin,
with $0.75 < z < 1.0$, contains 5666 galaxies; 42\% have spectroscopic
redshifts, 48\% have ANN$z$ photo-$z$'s, and 10\% have BPZ photo-$z$'s.
The third bin, with $0.75 < z < 1.0$, contains 5740 galaxies; 28\% have
spectroscopic redshifts, 45\% have ANN$z$ photo-$z$'s, and 27\% have BPZ
photo-$z$'s.  The sample statistics are summarized in Table
\ref{sample_table}.

\section{Determining Physical Properties}\label{methods}

The goal of this paper is to derive key physical quantities that can be
used to describe the stellar mass, evolutionary state, and environmental 
density of galaxies, and to use these measures to understand the physical 
processes that drive the broad patterns observed.  In this section we
discuss our methods for reliably determining these key variables by
making use of the unique combination of spectroscopy and near-IR imaging
offered by the DEEP2/Palomar survey described above.  The uncertainties
discussed below refer to the primary sample.  Our full error analysis is
described in $\S$\ref{uncertainties}.

\subsection{Stellar Masses Estimates}\label{masscode}

$K$-band luminosities alone provide stellar mass estimates that are
uncertain by factors of 5--7 \citep{brinchmann_thesis}.  However, for samples of known
spectroscopic redshift, using optical-infrared color information can 
constrain the stellar population and $M_*/L$ ratio so as to reduce this 
uncertainty to $\simeq$0.2 dex.  In this paper, we employ 
such a technique using the method described in \citet{bundy05} and 
discussed further in Bundy et al.~(in preparation).  We use a code 
that is Bayesian in nature and based on the precepts described 
in \citet{kauffmann03:mass} and \citet{BE00}.  

Briefly, the code uses $BRIK$ colors (measured using 2\farcs0 diameter
aperture photometry matched to the $K$-band detections) and
spectroscopic redshifts to compare the observed SED of a sample galaxy
to a grid of 13440 synthetic SEDs \citep[from][]{BC03} spanning a range
of star formation histories (parametrized as an exponential), ages
(restricted to be less than the age of the universe at the observed
redshift), metallicities, and dust content.  The choice of models and
population synthesis code employed can introduce systematic
uncertainties.  For example, \citet{maraston06} show that popular
recipes such as those in \citet{BC03} do not accurately chart the
evolution of Thermally-Pulsing Asymptotic Giant Branch stars.  For young
template models (with ages less than 2 Gyr), the Maraston models predict
stellar masses that can be lower by as much as $\sim$60\%.  We note this
potential source of error, but do not explicitly correct for it in
estimating stellar masses here.

Once the grid of models has been defined, the \kband~$M_*/L_K$, minimum
$\chi^2$, and the probability that each model accurately describes a
given galaxy is calculated at each grid point.  The corresponding
stellar mass is then determined by scaling $M_*/L_K$ ratios to the
\kband~luminosity based on the total \kband~magnitude and redshift of
the observed galaxy.  The probabilities are then summed (marginalized)
across the grid and binned by model stellar mass, yielding a stellar
mass probability distribution for each sample galaxy.  We use the median
of the distribution as the best estimate.

The stellar mass measured in this way is robust to degeneracies in the
model, such as those between age and metallicity. Although these
degeneracies can produce bimodal probability distributions, even in
these cases, the typical width of the distribution gives uncertainties
from the model fitting alone of 0.1--0.2 dex. For about 2--3\% of the
SED fits, the minimum $\chi^2$ values are significantly greater than
1.0.  For these more poorly-constrained objects, we add an additional
0.2 dex in quadrature to the final mass uncertainty. Although in
principle, the best fitting model also provides estimates of the age,
metallicity, star formation history, and dust content of a sample
galaxy, these quantities are much more affected by degeneracies and are
poorly constrained compared to the stellar mass.  

We also note that real galaxies are likely to have more complex star
formation histories than the simple stellar populations assumed here.
However, we do not attempt to fit the observed SEDs with more
complicated models, including those with multiple components and bursts,
because our photometric observations do not provide sufficient
constraints.  More complex models are particularly relevant to galaxies
with recent bursts of star formation which can hide underlying stellar
mass.  Several authors have investigated this effect.  Studying the most
extreme possibilities, \citet{shapley05:mass}, for example, find that
the stellar mass of blue galaxies with recent starbursts could be
underestimated by a factor of 5.  More typical values are probably much
less, and \citet{tran04} estimate the fraction of post starbursts
identified as E+A galaxies in the range $0.3 < z < 1$ to be a few
percent \citep[although see][]{yan05}, so the effect of starbursts in
our sample is likely to be small.

Photometric errors enter the stellar mass analysis by determining how
well the template SEDs can be constrained by the data.  Larger
photometric uncertainties ``smear out'' the portion of the model grid
that fits an observed galaxy with high probability.  This is reflected
in a wider stellar mass probability distribution.  Additional
uncertainties in the \kband~luminosity (from errors in the observed
total \kband~magnitude) lead to final stellar mass estimates that are
typically good to 0.2--0.3 dex.  The largest systematic source of error
comes from the assumed IMF, in this case that proposed by
\citet{chabrier03}.  The stellar masses we derive using this IMF can be
converted to Salpeter by adding $\sim$0.25 dex, although we note this
offset represents an average with a scatter across the models in our
grid of $\sim$0.06 dex.

\subsection{Indicators of Star Formation Activity}\label{U-B}

\begin{figure}
\plotone{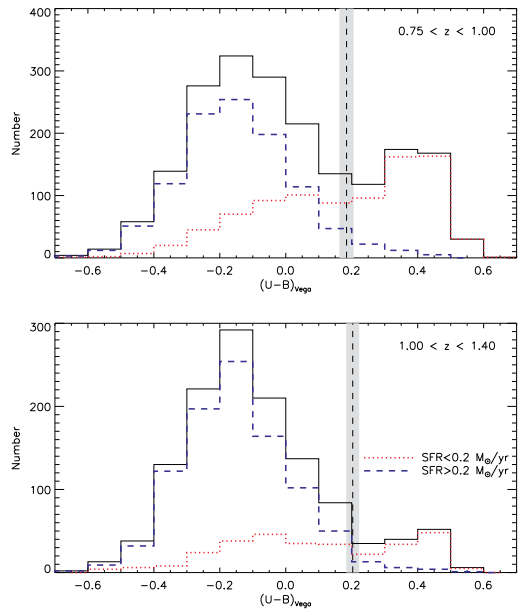}
\caption{The restframe $(U-B)$ color distribution in the middle and high
   redshift bins.  Overplotted are the distributions of galaxies whose star
    formation rate as derived from the [OII] equivalent width is greater than 
    0.2 \msun/yr (blue dashed line) and less  than
    0.2 \msun/yr (red dotted line).  The $(U-B)$ bimodality discriminant and
    its 1$\sigma$ range is indicated by the vertical dashed line and
   grey shading.\label{UB}}
\end{figure}

In this paper, we adopt two independent approaches for identifying those 
galaxies that are undergoing, or have recently experienced, active star formation.  
The first is the restframe $(U-B)$ color, estimated with the same methods as in
\citet{willmer05} and based on the {\sc imcat} photometry measured for
the CFHT $BRI$ data.  The $k$-corrections, which translate observed $BRI$
colors into restframe $(U-B)$ values, are determined by comparison to a
set of 43 nearby galaxy SEDs from \citet{kinney96}.  Second-order
polynomials were used to estimate $(U-B)$ and $k$-corrections as a
function of observed color.  Appendix A in \citet{willmer05} provides
more details on this technique.

The $(U-B)$ color distribution exhibits a clear bimodality that is used
to divide our sample into ``Blue'' (late-type) and ``Red'' (early-type)
subsamples using the same luminosity-dependent color cut employed by
\citet{vandokkum00}:

\begin{equation}\label{col_lim}
U - B = -0.032(M_B + 21.52) + 0.454 -0.25
\end{equation}

\noindent This formula, defined in the Vega magnitude system, divides
the sample well at all redshifts (see Figure \ref{UB}), and so we do not
apply a correction for potential rest-frame evolution with redshift
\citep{willmer05}.  We note that the proportion of red galaxies at
high-$z$ in Figure \ref{UB} is in part suppressed by the $R$-band
selection limit of $R_{AB} = 24.1$ (selection effects are discussed in
detail in $\S$\ref{completeness}).  

For galaxies with $z>0.75$, the [OII] 3727 \AA\ emission line falls
within the spectroscopic range probed by the DEEP2 Survey and this can
be used to provide a second, independent estimate of the SFR.  For
these galaxies, we measure the intensity of the [OII] emission line by
fitting a double gaussian---with the wavelength ratio constrained to
that of the [OII] doublet---using a non-linear least squares
minimization.  We measure a robust estimate of the continuum by taking
the biweight of the spectra in two wavelength windows each 80 \AA\ long
and separated from the emission line by 20 \AA.  We consider only
spectra where the equivalent width of the feature is detected with
greater than 3$\sigma$ confidence.

Because the DEEP2 spectra are not flux calibrated, we use the following 
formula from \citet{guzman97} to estimate the [OII] SFR:

\begin{equation}\label{eq_SFR}
{\rm SFR(M_{\odot}~yr^{-1})} \approx 10^{-11.6 -0.4(M_B -
   M_{B\odot})}{\rm EW_{[OII]}}
\end{equation}

\noindent This relation utilizes the restframe $M_B$ estimated in the
same way as the $(U-B)$ colors \citep[see][]{willmer05} and provides an
approximate value for the SFR without correcting for metallicity effects,
which can introduce random and systematic errors of at least 0.3--0.5
dex in the SFR derived from ${\rm L_{[OII]}}$ \citep{kewley04}.  In
addition, Equation \ref{eq_SFR} was optimized for a sample of compact
emission line galaxies \citep[see][]{guzman97} which likely differs in
the amount of extinction and typical [OII]/H$\alpha$ ratio compared to
the sample here, yielding SFR estimates that could be systematically off
by a factor of $\sim$3.  Moreover, recent work by \citet{papovich05}
demonstrates that SFR's based on re-radiated IR radiation can be orders
of magnitude larger than optical/UV estimates, and studies of the AGN
population in DEEP2 indicate that [OII] emission is often associated
with AGN, further biasing [OII]-based SFR estimates \citep{yan05}.
Future efforts, especially those utilizing the multi-wavelength data set
in the EGS, will be useful in refining the SFR estimates.  In the
present paper, we accept the limitations of Equation \ref{eq_SFR} for
determining absolute SFR's, noting that our primary requirement is not a
precision estimate of the SFR for each galaxy but only the broad
division of the field population into active and quiescent subsets.

In support of this last point, Figure \ref{UB} compares the $(U-B)$ and
[OII] star formation indicators directly in the two higher redshift bins
where both diagnostics are available.  The solid histogram traces the
full $(U-B)$ distribution with the vertical dashed line (and shading)
indicating the median and 1$\sigma$ range of the color bimodality
discriminant (the range results from the dependence on $M_B$ in Equation
\ref{col_lim}).  Using the independent diagnostic of the [OII]
equivalent width, we can similarly divide the population into high
(blue dashed histogram) and low (red dotted histogram) star-forming
populations using a cut of ${\rm SFR_{[OII]}}= 0.2$ \msun~yr$^{-1}$,
which is the median SFR of galaxies with $0.75 < z < 1.0$.  Figure
\ref{UB} clearly shows the effectiveness of the $(U-B)$ color cut in
distinguishing the populations in both cases.  The fraction of red
galaxies in the high ${\rm SFR_{[OII]}}$ population is less than 8\% in
the middle redshift bin and less than 17\% in the high-$z$ bin.  The
fact that a non-zero fraction of blue galaxies is made up of the low
${\rm SFR_{[OII]}}$ population is likely indicative of the 1--2 Gyr
timescale required for galaxies to join the red sequence and suggests
that even minor episodes of star formation can lead to blue restframe
colors \citep{gebhardt03}.  In addition, the median value of the
measured [OII] SFR increases in the high-$z$ bin.  This means that
galaxies satisfying the ${\rm SFR_{[OII]}}= 0.2$ \msun~yr$^{-1}$ cut at
high redshift are more vigorously forming stars, and therefore bluer.

Further details on the evolving SFR will be presented in Noeske et
al.~(in preparation).  We also note that the $(U-B)$ and [OII] star formation
indicators are consistent with the star formation histories recovered by
the SED template fitting procedure used to refine stellar mass estimates
and described in $\S$\ref{masscode}.  This agreement is expected because
the restframe color and SED fitting are both determined by the observed
colors.  Such consistency demonstrates that the measured physical
properties that we use to divide the galaxy population are also
reflected in the best-fit SED templates that determine stellar mass (see
$\S$\ref{masscode}).

\subsection{Environmental Density}\label{density}

\begin{figure}
\plotone{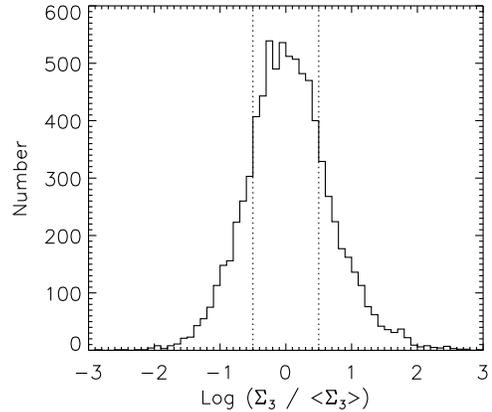}
\caption{The distribution in the relative overdensity in log units as
   measured across the sample by the 3$^{rd}$-nearest-neighbor  
   statistic as implemented by \citet{cooper05}. The vertical dotted lines 
    at $\pm$0.5 dex divide the distribution into low, middle, and high 
    density regimes (see text for details).\label{dens_dist}}
\end{figure}

Charting galaxy evolution over a range of environments represents a key
step forward that can only be achieved through large spectroscopic
redshift surveys such as in the DEEP2 survey.  \citet{cooper05}
rigorously investigate the question of how to provide precise
environmental density estimates in the context of redshift surveys at $z
\sim 1$. That work clearly shows, via comparisons to simulated samples,
that large samples with photo-$z$'s are very poorly suited to providing
accurate density measures in all but the most underdense environments.

Specifically, \citet{cooper05} find that for environmental studies at high-$z$,
the projected $n^{th}$-nearest-neighbor distance, $D_{p,n}$, offers the
highest accuracy over the greatest range of environments.  This measure
is the field counterpart to projected density estimates first applied to
studies of cluster environments \citep[e.g.,][]{dressler80}.  The
statistic is defined within a velocity window ($\Delta v = \pm 1000$ km
s$^{-1}$) used to exclude contaminating foreground and background
galaxies.  It is therefore particularly robust to redshift space
distortions in high density environments without suffering in accuracy
in underdense environments.  In addition, the effect of survey
boundaries on $D_{p,n}$ is easily understood and mitigated by excluding
a small strip around survey edges.

In this paper, we utilize the projected $3^{rd}$-nearest-neighbor
distance, $D_{p,3}$, excluding galaxies closer than 1 $h^{-1}$ Mpc
($\sim$3\arcmin) from a survey boundary.  The choice of $n = 3$ does not
significantly affect $D_{p,n}$ which has a weak dependence on $n$ in
both high and low density environments for $n \leq 5$ \citep{cooper05}.
The effects of survey target selection must be carefully considered
because they can introduce biases as a function of redshift.
\citet{cooper05} find that the sampling rate and selection function in
the DEEP2 survey equally probes all environments at $z \sim 1$ in a
uniform fashion.  Although the DEEP2 survey secures redshifts for
roughly 50\% of galaxies at $z\sim 1$, its sparse selection algorithm
does not introduce a significant environment-dependent bias. However,
the absolute value of $D_{p,3}$ will increasingly underestimate the true
density with increasing redshift as the sampling rate declines.  To
handle this effect, we first convert $D_{p,3}$ into a projected surface
density, $\Sigma_3$, using $\Sigma_3 = 3 / (\pi D_{p,3}^2)$.  We then
calculate the {\em relative} overdensity at the location of each galaxy
by dividing the observed value of $\Sigma_3$ by the median surface
density calculated in bins of $\Delta z = 0.04$.  The relative
overdensity is thus insensitive to the DEEP2 sampling rate, providing a
reliable statistic that can be compared across the full redshift range
of the sample \citep{cooper05}.  We note that a small bias in this
statistic may be introduced by the fact that DEEP2 preferentially
selects blue galaxies as the redshift increases.  Tests indicate this
effect is smaller than the typical density uncertainty (a factor of
$\sim$3) and is not likely to be important given the coarse density bins
we use to divide the sample.  Further details are provided in
\citet{cooper06}.

The distribution of the relative overdensity for our primary sample is 
plotted in Figure \ref{dens_dist}.  In the analysis that follows, we 
consider two ways of dividing the sample by environmental density.  
In the first, we separate galaxies according to whether they lie in
regions above or below the median density (corresponding to a 
measured overdensity of zero in Figure \ref{dens_dist}). This is
the simplest criterion but does mean the bulk of the signal
is coming from regions which are not too dissimilar in their environs. 
In the second approach, we divide the density sample into three bins as
indicated by the vertical dotted lines in Figure \ref{dens_dist}.  The
thresholds of 0.5 dex, or 0.77$\sigma$, above and below the median
density were chosen to select the extreme ends of the density
distribution where the field sample begins to probe group and
void-like environments that are 3--100 times more or less dense than
average. The conclusions presented in $\S$\ref{results} are not
sensitive to the precise location of these three thresholds.

\section{Constructing the Galaxy Stellar Mass Function}\label 
{mfn_description}

The DEEP2/Palomar survey presents a unique data set for constraining the
galaxy stellar mass function at $z \sim 1$.  Previous efforts have so
far relied on smaller and more limited samples, often without
spectroscopic redshifts.  Only via the combination of size,
spectroscopic completeness, and depth, does the current sample enable us
to study the evolving relationships between stellar mass, star formation
activity, and local environment in a statistically robust way.  Our
primary tool in this effort is charting the galaxy stellar mass function
of various populations.

Deriving the stellar mass function in a magnitude limited survey requires
corrections for the fact that faint galaxies are not detected throughout
the entire survey volume.  The $V_{max}$ formalism \citep{schmidt68} is
the simplest technique for handling this problem but does not account
for density inhomogeneities that can bias the shape of the derived mass
function.  While other techniques address this problem \citep[for a
review, see][]{willmer97}, they face different challenges such as determining
the total normalization.  For sufficiently large samples over
significant cosmological volumes, density inhomogeneities cancel out and
the $V_{max}$ method produces reliable results.  Considering this and
the fact that we wish to use our data set to explicitly test for the
effects of density inhomogeneities, we adopt the $V_{max}$ approach, which
offers a simple way to account for both the $R$ and \kband~limits of our
sample.  For each galaxy $i$ in the redshift interval $j$, the value of
$V^i_{max}$ is given by the minimum redshift at which the
galaxy would leave the sample, becoming too faint for either the $R$ or
\kband~limit.  Formally, we define

\begin{equation}
V^i_{max} = d \Omega_j \int_{z_{low}}^{z_{high}} \frac {dV}{dz} dz
\end{equation}

\noindent where $d \Omega_j$ is the solid angle subtended by the sample
defined by the limiting \kband~magnitude, $K^j_{lim}$, for the redshift
interval $j$, and $dV/dz$ is the comoving volume element.  The redshift
limits of the integral are:

\begin{equation}
z_{high} = {\rm min}(z^j_{max}, z^j_{K_{lim}}, z_{R_{lim}})
\end{equation}
\begin{equation}
z_{low} = z^j_{min}
\end{equation}

\noindent where the redshift interval, $j$, is defined by $[z^j_{min},
z^j_{max}]$, $z^j_{K_{lim}}$ refers to the redshift at which the galaxy
would still be detected below the \kband~limit for that particular
redshift interval, and $z_{R_{lim}}$ is the redshift at which the galaxy
would no longer satisfy the $R$-band limit of $R_{AB} \leq 24.1$.  We use the
SED template fits found by the stellar mass estimator to calculate
$z^j_{K_{lim}}$ and $z_{R_{lim}}$, thereby accounting for the
$k$-corrections necessary to compute accurate $V_{max}$ values (no
evolutionary correction is applied).

In constructing the $V_{max}$ mass functions, we also weight the
spectroscopic sample to account for incompleteness resulting from the
DEEP2 color selection and redshift success rate.  We closely follow the
method described in \citet{willmer05}, but add an extra dimension to the
reference data cube which stores the number of objects with a given
\kband~magnitude.  Thus, for each galaxy $i$ we count the number of
objects from the photometric catalog sharing the same bin in the
$(B-R)$/$(R-I)$/$R_{AB}$/\kband~parameter space as well as the fraction
of these with attempted and successful redshifts.  As mentioned in
$\S$\ref{DEEP2}, $\approx$30\% of attempted DEEP2 redshifts are
unsuccessful, mostly because of faint, blue galaxies beyond the redshift
limit accessible to DEEP2 spectroscopy \citep{willmer05}.  Failed
redshifts for red galaxies are more likely to be within the accessible
redshift range but simply lacking in strong, identifiable features.  We
therefore use the ``optimal'' weighting model \citep{willmer05}, which
accounts for the redshift success rate by assuming that failed redshifts
of red galaxies (defined by the $(U-B)$ color bimodality) follow the
same distribution as successful ones, while blue galaxies with failed
redshifts lie beyond the redshift limit ($z \approx 1.5$) of the sample.
The final weights are then based on the probability that a successful
redshift would be obtained for a given galaxy.  They also account for
the selection function applied to the EGS sample to balance the fraction
of redshifts above and below $z \approx 0.7$.  With the weight,
$\chi_i$, calculated in this way, we determine the differential galaxy
stellar mass function:

\begin{equation}
\Phi(M_*) dM_* = \sum_i \frac {\chi_i}{V^i_{max}} dM_*
\end{equation}

\subsection{Uncertainties and Cosmic Variance}\label{uncertainties}

\begin{figure*}
\plotone{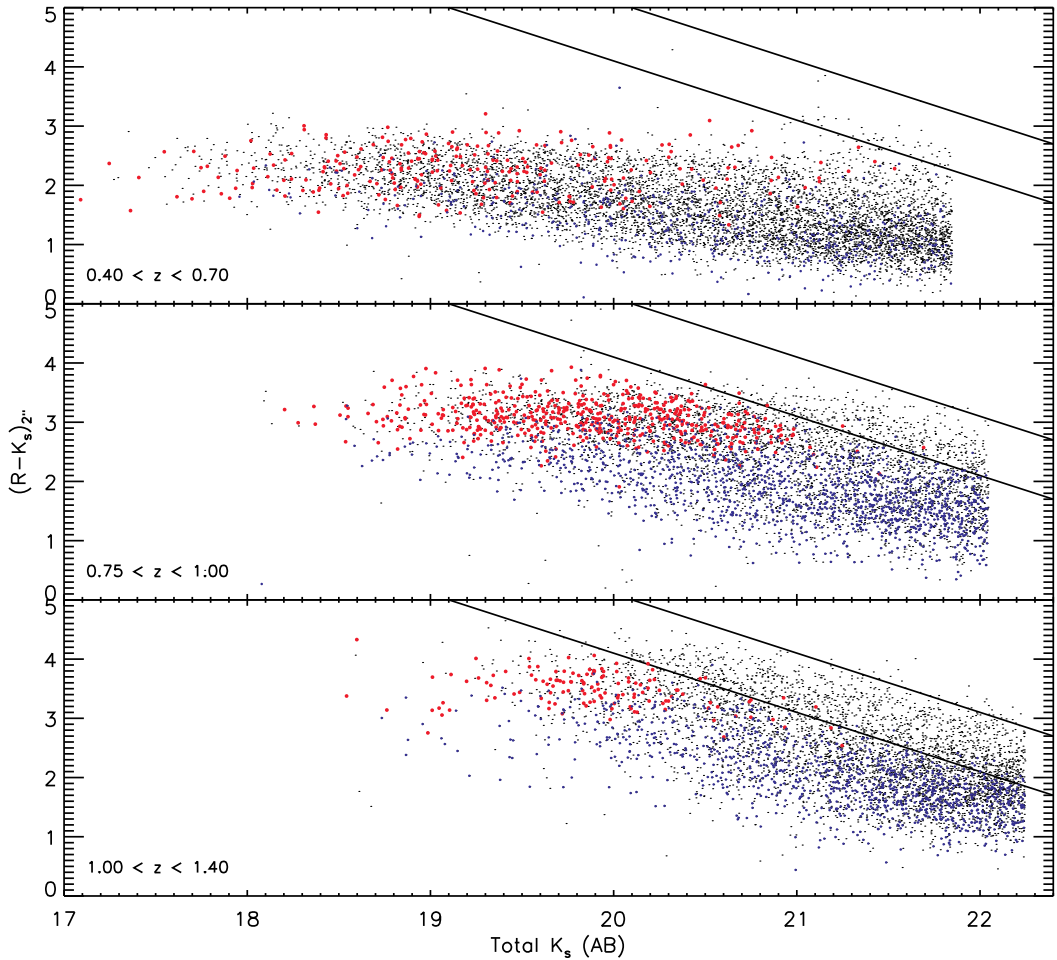}
\caption{Apparent color-magnitude diagrams used to illustrate sample completeness.  The
   plots show the distribution of $(R-K_s)$ vs $K_s$ for the
   primary spectroscopic redshift sample with $R_{AB} \leq 24.1$ (solid, color  circles)
   compared to the $R_{AB} \leq 25.1$ sample, which has been  supplemented with
   photometric redshifts (small black dots).  The spectroscopic sample is
   colored according to location in the bimodal restframe $(U-B)$ distribution as
   described in the text.  The $R_{AB} = 24.1$ and $R_{AB} = 25.1$  
   magnitude limits  are indicated by the solid diagonal lines.   \label{RK}}
\end{figure*}

\begin{figure*}
\plotone{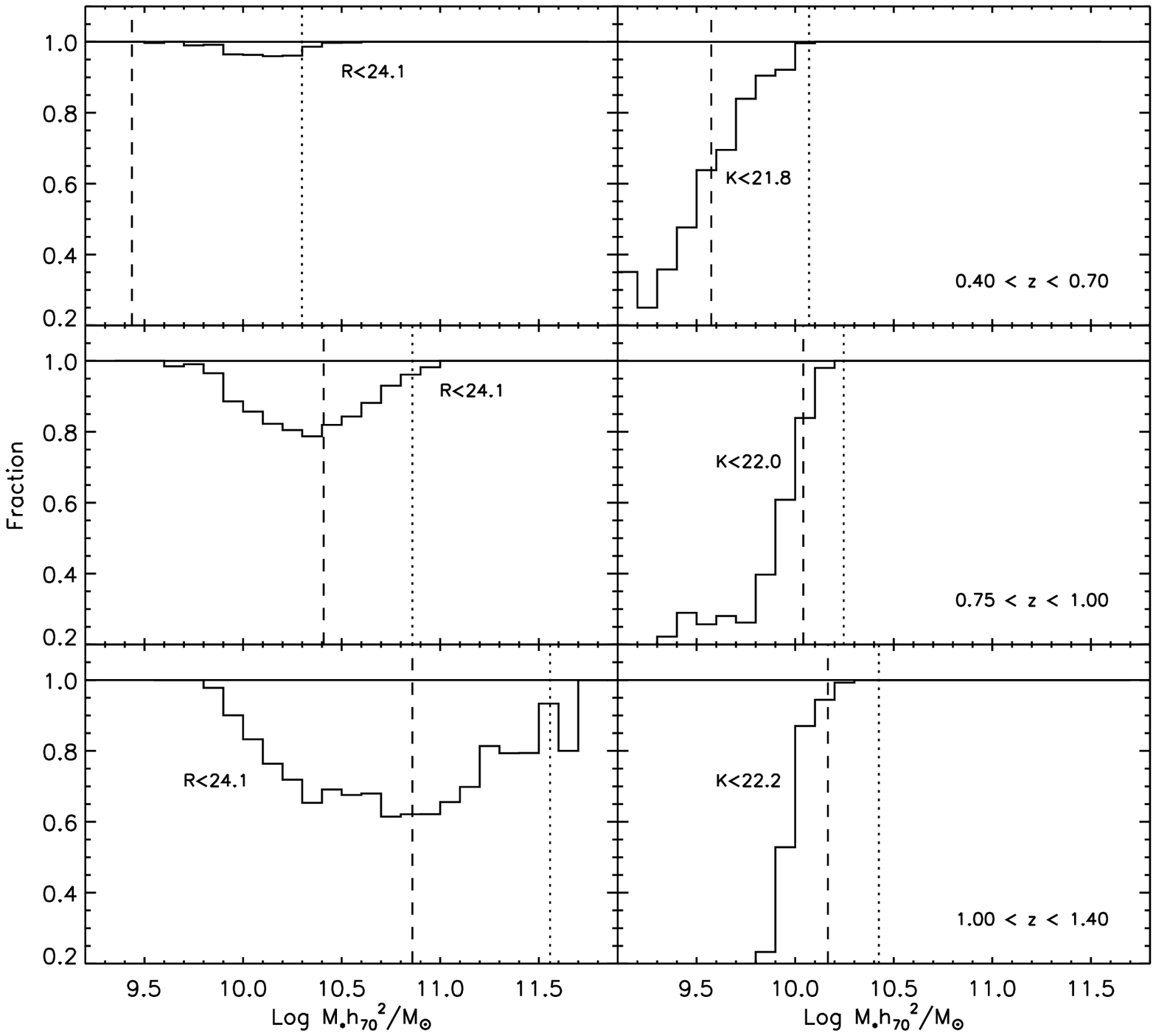}
\caption{Completeness of the mass distribution in the primary
  spectroscopic sample taking into account the $R$-band and
  \kband~magnitude cuts.  The left-hand panels show the completeness
  resulting from the $R_{AB} \leq 24.1$ limit as derived from
  comparisons within the photo-$z$ supplemented sample.  Vertical dotted
  lines indicate the mass of the faintest galaxy that could be detected
  at the high-$z$ end of each redshift interval, while vertical dashed
  lines correspond to galaxies at the low-$z$ end.  The right-hand
  panels, also drawn from the $R_{AB} \leq 25.1$ sample, show the effect
  of the \kband~limit on the mass completeness.  Vertical dotted and
  dashed lines indicate model galaxy limits, as in the left panels.  The
  more conservative dotted lines are adopted as the mass limits of the
  paper.\label{mass_complete}}
\end{figure*}

To estimate the uncertainty in our mass distribution we must account for several
sources of random error and model their combined effect through Monte
Carlo simulations.  Different error budgets are calculated for the primary
spectroscopic and photo-$z$ supplemented samples via 1000
realizations of our data set in which we randomized the expected errors.  In
both samples we model uncertainties in $V_{max}$ arising from
photometric errors, and simulate the error on the stellar mass
estimates, which, as described in $\S$\ref{masscode}, is typically 0.2
dex and is encoded in the stellar mass probability distribution of each
galaxy.

For the primary spectroscopic sample, errors on the rest-frame $(U-B)$
colors are estimated by noting the photometric errors for a given
galaxy.  We do not model the uncertainty in the [OII] SFR because 
unaccounted systematic errors are likely to be greater than the random
uncertainty of measurements of the [OII] equivalent width. We stress again
that these diagnostics are only used to separate the bimodal distribution
into active and quiescent components.

For the photo-$z$ supplemented sample, we model redshift uncertainties
using $\Delta z / (1+z) = 0.07$ for the $R_{AB} \leq 24.1$ ANN$z$
subsample and $\Delta z / (1+z) = 0.18$ for the $R_{AB} \leq 25.1$ BPZ
subsample.  The BPZ uncertainty is slightly higher than measured in the
comparison to the spectroscopic sample because we expect a poorer
precision of BPZ on objects with $R_{AB} > 24.1$.  These redshift
uncertainties affect the distribution of objects in our redshift bins as
well as the Monte Carlo realizations of stellar mass estimates in the
photo-$z$ sample. The final uncertainty at each data point in the
stellar mass functions from both the spectroscopic and photo-$z$
supplemented samples is estimated conservatively as the sum, in
quadrature, of the 1$\sigma$ Monte Carlo errors and the Poisson errors.

We now turn to cosmic variance.  Because our sample is drawn from four
independent fields, it is possible to estimate the effects of cosmic
variance by comparing the results from different fields in each of our
three redshift bins: $0.4 < z < 0.7$, $0.75 < z < 1.0$, and $1.0 < z <
1.4$.  For the spectroscopic sample with $z > 0.75$, we can compare the
EGS to the sum of Fields 2, 3, and 4, yielding two subsamples with
roughly equal numbers of galaxies.  We compare the total and
color-dependent stellar mass functions of these two subsamples and
divide the median of the differences measured at all data points by
$\sqrt 2$ to estimate the cosmic variance.  Unfortunately, galaxies with
$z < 0.75$ come only from the EGS.  The cosmic variance estimate here is
derived by performing the same calculation on three subsets of the EGS
sample and dividing it by $\sqrt 3$. Of course, cosmic variance on the
scale of the EGS itself is not included in this estimate.  For the three
redshift intervals, this method provides {\em average} 1$\sigma$
systematic cosmic variance uncertainties of 29\%, 12\%, and 26\%.

In addition to these rough estimates, we have checked that the observed
density distribution is not affected by cosmic variance.  Based on the
density-dependent mass functions from different fields, there is no
evidence of a single structure or overdensity in one of the fields that
would bias our results.  We also note that the type-dependent mass
functions are, to first order, affected by cosmic variance in the same
way as the total mass functions.  Thus, while absolute comparisons
between different redshift intervals must account for cosmic variance
errors, comparisons using the relative or fractional abundance of a
given population are less susceptible to this uncertainty.  For example,
comparisons between different fields of the abundance of blue galaxies
at $z \sim 0.5$ with masses near $M_* = 10^{10.8}$\msun\ (see
$\S$\ref{results}) suggest that the absolute measurement is uncertain at
the $\sim$40\% level, while the relative measurement made with respect
to the total abundance of galaxies is uncertain only at the $\sim$15\%
level.  For red galaxies in this same bin, the standard deviation for
the relative fraction is 10\%, but increases to 25\% for the measurement
of their absolute abundance.

\subsection{Completeness and Selection Effects}\label{completeness}

We now turn to the important question of the redshift-dependent
completeness of the stellar mass functions in our sample.
Incompleteness resulting from the DEEP2 selection technique is corrected
for by applying weights to galaxies within the spectroscopic sample, as
described in $\S$\ref{mfn_description}.  The $R$ and \kband\ limits of
this sample also introduce incompleteness effects, however, that cannot
be corrected through weighting.  Instead, we adopt two approaches to
determine how the incompleteness from our magnitude limits affects the
derived mass functions.

First, we consider the likelihood of observing model galaxies of known
mass and color based on the $R$ and \kband~magnitude limits of the
primary spectroscopic sample.  We follow previous work
\citep[e.g.,][]{fontana03} and track the stellar mass of a template
galaxy with a reasonable maximum $M_*/L$ ratio determined by models with
solar metallicity, no dust, and a burst of star formation beginning at
$z_{form} = 5$ that lasts for 0.5 Gyr and is then truncated.  We set the
luminosity according to the $R$ and \kband~limiting magnitudes.  The
maximum stellar masses of these model galaxies provide estimates of our
completeness limits.  We consider two realizations.  In the first, the
model is placed at the low-$z$ edge of each redshift interval where the
intrinsically faintest galaxies would be detected (such galaxies have
high $V_{max}$ weights).  In the second, more conservative limit, the
model is placed at the high-$z$ edge of the interval, representing the
faintest systems that could be detected across the entire redshift bin.
The results indicate that the \kband\ limit ensures completeness above
$3 \times 10^{10}$\msun\ at all redshifts, while the $R$-band limit
primarily affects our high-$z$ sample below $10^{11}$\msun.  We discuss
these limits in detail below.

Our second approach is to measure the effects of incompleteness  
directly by comparing our primary spectroscopic sample with $R_{AB} \leq 24.1$ to the  
fainter sample with $R_{AB} \leq 25.1$, supplemented by photometric redshifts.   
This approach is particularly useful for investigating the way in which the
$R$-band limit introduces a bias against red galaxies, especially at $z
\gtrsim 1$.  This bias could mimic the effect of downsizing by suppressing
the fraction of red galaxies at $z \gtrsim 1$.  The behavior of model
galaxies (described above) as well as the comparison to a fainter sample
both yield consistent estimates for the sample completeness which we  
will show does not compromise our results.

Figure \ref{RK} compares the distribution in 2\farcs0 diameter $(R-K_s)$
vs total $K_s$ color-magnitude space of the primary spectroscopic sample
with $R_{AB} \leq 24.1$ (solid color circles) to that for the fainter
$R_{AB} \leq 25.1$ sample (small black dots) supplemented with
photometric redshifts. As expected, in the low redshift bin, the
majority of the $R_{AB} \leq 25.1$ sample is contained within the
spectroscopic limit of $R_{AB} = 24.1$ (although the photo-$z$ sample
includes many more galaxies with $R_{AB} \leq 24.1$ that were not
selected for spectroscopy).  At high redshift, however, the primary
sample is clearly incomplete, with a substantial number of galaxies
having $R_{AB} > 24.1$.  While the full range of $(R-K_s)$ colors is
included in the primary sample, a color bias is introduced because the
reddest galaxies are no longer detected at $K_{AB} \gtrsim 20$.

To demonstrate how this color bias affects the mass completeness of the
sample, the left-hand panels in Figure \ref{mass_complete} chart the
stellar mass completeness of the photo-$z$ supplemented sample limited
to $R_{AB} \leq 24.1$  as compared to the full sample
with $R_{AB} \leq 25.1$.  With increasing redshift,
the fraction of galaxies satisfying $R_{AB} \leq 24.1$ decreases.  This
results in a loss of detected galaxies below a given mass which
corresponds closely to the conservative limits based on model galaxies
placed at the high-$z$ edge of each redshift interval, as described
above and indicated by the dotted vertical lines in Figure
\ref{mass_complete}.  However, because of the $V_{max}$ weighting, our
analysis is actually more complete than the figure suggests.  A more
appropriate completeness limit is indicated by the vertical dashed line,
which corresponds to model galaxies at the low-$z$ edge of each redshift
interval where the faintest galaxies are still detected.

Figure \ref{mass_complete} indicates that for galaxies with $z < 1$,
the $R$-band limit has only a small effect on the spectroscopic sample.
Galaxies with $R_{AB} \leq 24.1$ constitute the vast majority of the
total distribution in the first two redshift bins, with a minimum
completeness of $\approx$80\% in the middle redshift bin.  The $R$-band
limit has a greater effect in the high-$z$ bin, where the $R_{AB} \leq
24.1$ distribution accounts for just over $\approx$60\% of the $R_{AB}
\leq 25.1$ sample over most of the mass range.  We note that the
$R$-band limit affects mostly red galaxies which we will show dominate
at higher masses.  For this reason, the completeness function resulting
the $R$-band limit rises again at lower masses.  In section
$\S$\ref{results}, we correct for the $R$-band incompleteness and show
that it does not affect our conclusions.

The right-hand panels of Figure \ref{mass_complete} illustrate the
effect of the \kband~limit on the mass completeness of the spectroscopic
sample.  Using our deepest Palomar observations, we construct stellar
mass distributions from the photo-$z$ supplemented sample with $R_{AB}
\leq 25.1$.  Again, for each redshift interval, we compare the mass
distribution of a subsample with a \kband~limit equal to 0.5 magnitudes
fainter than the limit imposed on the spectroscopic sample at that
redshift, $K^j_{lim}$.  As in the left-hand panels of Figure
\ref{mass_complete}, the resulting completeness function yields an
estimate of the \kband~mass completeness limit that is consistent with
the high-$z$ and low-$z$ model galaxy masses described above.  For the
analysis to follow, we conservatively adopt the high-$z$ model limits
indicated by the dotted lines, although we acknowledge that heavily
obscured galaxies may still be missed above these limits.  For each
redshift bin, these limits correspond to 10.1, 10.2, and 10.4 in units
of $\log$ \msun.  We note that for the middle and high redshift bins in
which the $R$-band limit affects the sample, the \kband\ completeness
limit occurs at lower masses, justifying our method for studying the $R$
and \kband\ limits separately.

\section{Results}\label{results}

\subsection{The Total Stellar Mass Function}

\begin{figure}
\plotone{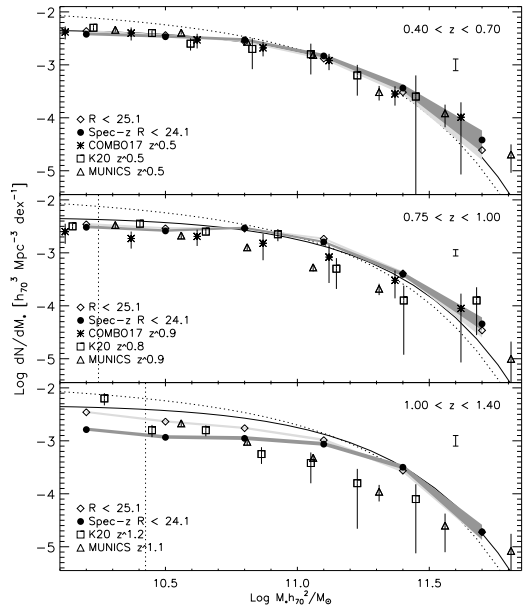}
\caption{Total stellar mass functions in three redshift bins.  Shading
  indicates the width of 1$\sigma$ error bars. The mass function for all
  galaxies in the primary spectroscopic sample is designated by solid
  circles with dark shading. That for the deeper, $R_{AB} \leq 25.1$
  sample, supplemented by photo-$z$'s, is indicated by open diamonds and
  light gray shading.  Results from the COMBO17 \citep{borch06}, K20
  \citep{fontana04}, and MUNICS \citep{drory04} surveys are also
  plotted.  The dotted curve is the Schechter fit to the local mass
  function from \citet{cole01}.  The solid curve is a Schechter fit to
  the $R_{AB} \leq 25.1$ results in the first redshift bin.  Vertical
  dotted lines show estimates for the mass completeness limit resulting from the
  \kband~magnitude limits---in the highest redshift interval
  incompleteness resulting from the $R$-band limit becomes important, as
  demonstrated by the difference between the spectroscopic and photo-$z$
  supplemented mass functions.  The isolated error bar in the upper
  right-hand portion of each plot indicates the estimated average
  uncertainty due to cosmic variance.
\label{mfn_tot_ref}}
\end{figure}

The total galaxy stellar mass functions in three redshift intervals are
plotted in Figure \ref{mfn_tot_ref}.  Results for both the
spectroscopic, $R_{AB} \leq 24.1$ sample (solid circles, dark shading)
and $R_{AB} \leq 25.1$, photo-$z$ supplemented sample (open diamonds,
light shading) are presented.  The width of the shaded curves
corresponds to the final 1$\sigma$ errors using the Monte Carlo
techniques discussed earlier.  For the $R_{AB} \leq 24.1$ sample, the
lowest redshift interval draws only from the EGS, leading to a larger
cosmic variance uncertainty as indicated by the isolated error bar in
the upper right portion of the plot.  The $R_{AB} \leq 24.1$ mass
functions in the two higher redshift bins, as well as the $R_{AB} \leq 25.1$
sample in all bins, utilize data from all four DEEP2 survey fields.  The
close agreement between the spectroscopic and photo-$z$ supplemented
total mass functions indicates that the weighting applied to compensate
for the DEEP2 spectroscopic selection function in the $R_{AB} \leq 24.1$
sample is successful at recovering a complete, magnitude-limited sample.
The vertical dotted lines represent estimates of the mass completeness
originating from the \kband~magnitude limit (see
$\S$\ref{completeness}).  The solid curve plotted for all three redshift
intervals is the Schechter fit to the $R_{AB} \leq 25.1$ mass function
at $z \sim 0.55$, and can be used to gauge evolution within our sample.

As discussed in $\S$\ref{completeness}, the mass functions of the
primary sample presented in Figure \ref{mfn_tot_ref} are affected by
incompleteness because of the $R_{AB} = 24.1$ spectroscopic limit.  The
degree of incompleteness is apparent in the comparison between the total
mass functions of the spectroscopic and $R_{AB} \leq 25.1$ samples.  As
expected from Figure \ref{mass_complete}, the first two redshift bins
are hardly affected by $R$-band incompleteness.  However, the $R_{AB} =
24.1$ limit is clearly important in the high redshift bin where the
comoving number density in the photo-$z$ sample is larger by a factor of
$\approx$2 (0.3 dex) for $M_* < 10^{10.8}$\msun.

Figure \ref{mfn_tot_ref} also plots results from several previous
studies, all of which have been normalized to $h=0.7$ and adjusted for
comparison to the Chabrier IMF used here.  The dotted curve is the
Schechter fit to the local stellar mass function as determined by
\citet{cole01}.  Results from the K20 \citep{fontana04}, MUNICS
\citep{drory04}, and COMBO17 \citep{borch06} surveys are also plotted.
The K20 survey \citep{cimatti02} has good spectroscopic coverage (92\%)
but is shallower than the data presented here and covers only 52
square arcminutes or 0.01 square degrees.  The MUNICS survey \citep{drory01}
covers roughly one square degree, but only reaches $K \sim 19.5$ (Vega)
and is composed of 90\% photometric redshifts.  The COMBO17 results
utilize carefully calibrated photo-$z$'s \citep{wolf04} and represent an
area of 0.8 square degrees, but infrared photometry was not available.
Instead, \citet{borch06} scale their stellar mass estimates to the
magnitude measured in a narrow filter (21 nm wide) at 816\AA.  This
restricts the redshift range where masses can be estimated to $z < 1$.

The stellar mass functions from previous work as well as the new results
presented here are in good agreement in the first redshift bin in Figure
\ref{mfn_tot_ref}.  It is interesting to note that all of the results
plotted find a higher abundance of massive galaxies ($M_* \gtrsim 5
\times 10^{11}$\msun) and a lower abundance of less massive galaxies
($M_* \lesssim 6 \times 10^{10}$\msun) compared to the local mass
functions of \citet{cole01}, although a slight excess above the
Schechter fit is also seen at the highest masses in local samples
\citep[e.g.,][]{cole01}.  We do not plot the local mass function from
\citet{bell03:LF} because it is in good agreement with \citet{cole01}.
\citet{borch06} argue that the higher number of massive galaxies in
their study may be caused by photo-$z$ errors, but we find similar
results in our spectroscopic sample.  This discrepancy may reflect
systematic differences in the way stellar masses are estimated, and
highlights potential difficulties in comparing high-$z$ work to local
studies.

Our mass function at $z \sim 0.9$ shows little evolution compared to the
lower redshift interval.  In fact, for $M_* \gtrsim 10^{11}$\msun, we
find slightly higher number densities at $z \sim 0.9$, but this is
likely due to cosmic variance.  Our results are consistent with the
COMBO17 mass functions but are higher than the K20 and MUNICS mass
functions, a difference that seems to increase in the highest redshift
bin.  Here we find only a slight decrease in the numbers of galaxies
with $M_* \gtrsim 3 \times 10^{11}$\msun, an effect that is consistent
with no evolution, given the uncertainties from cosmic variance.  The
photo-$z$ supplemented mass function at $z \sim 1.2$ reveals a more
significant decline, however, with respect to our low-$z$ bin in the
number density of lower mass galaxies (with $M_* \lesssim 3 \times
10^{11}$\msun).  This hints at the notion that the mass assembly history
of all galaxies may proceeds in a top-down fashion, lending support to
recent claims of so-called ``mass assembly downsizing''
\citep[e.g.,][]{cimatti06}.  While intriguing, we postpone a detailed
analysis of this effect to a future paper, choosing instead to focus on
downsizing in the context of star formation activity.

\begin{deluxetable}{lccc}
\tablecaption{Schechter Fit Parameters}

\tablewidth{0pt}
\tablecolumns{4}
\tablehead{
\colhead{} & \colhead{$\phi^*$} & \colhead{$\log h_{70} M^*$} &
\colhead{$\alpha$} \\
\colhead{Redshift} & \colhead{($h_{70}^3$ Mpc$^{-3}$ dex$^{-1}$)} & \colhead{(\msun)} &
\colhead{} \\
\cline{1-4} \\
\multicolumn{4}{c}{Primary Spec-$z$, $R_{AB} \leq 24.1$} 
}


\startdata
$0.4 < z < 0.7$  & $0.0027 \pm 0.0004$ & $10.94 \pm 0.08$ & $-0.81 \pm
0.2$ \\
$0.75 < z < 1.0$ & $0.0031 \pm 0.0010$ & $10.87 \pm 0.10$ & $-0.59 \pm
0.4$ \\
$1.0 < z < 1.4$  & $0.0012 \pm 0.0008$ & $10.97 \pm 0.25$ & $-0.51 \pm
0.5$ \\
\cutinhead{Photo-$z$ Supplemented, $R_{AB} \leq 25.1$}
$0.4 < z < 0.7$  & $0.0025 \pm 0.0007$ & $10.93 \pm 0.11$ & $-0.95 \pm
0.1$ \\
$0.75 < z < 1.0$ & $0.0031 \pm 0.0014$ & $10.87 \pm 0.10$ & $-0.51 \pm
0.4$ \\
$1.0 < z < 1.4$  & $0.0015 \pm 0.0004$ & $11.01 \pm 0.07$ & $-0.93 \pm
0.1$ \\

\enddata
\label{sch_fit}
\tablecomments{Quoted uncertainties were estimated based on
  field-to-field comparisons.}
\end{deluxetable}

In Table \ref{sch_fit}, we list the parameters of Schechter fits to the
$R_{AB} \leq 24.1$ and photo-$z$ supplemented, $R_{AB} \leq 25.1$ mass
functions plotted in Figure \ref{mfn_tot_ref}.  By integrating the
photo-$z$ Schechter fits we obtain estimates for the global stellar mass
density in units of $\log$ \msun\ $h_{70}$ Mpc$^{-3}$ of $\log \rho_* =
8.31 \pm 0.07$ at $z \sim 0.55$, $\log \rho_* = 8.30 \pm 0.1$ at $z \sim
0.9$, and $\log \rho_* = 8.15 \pm 0.1$ at $z \sim 1.2$, where the
uncertainties come from cosmic variance estimates based on
field-to-field comparisons.  These results are in good agreement with
previous measurements of $\rho_*$ \citep{dickinson03, rudnick03, drory05}.

\subsection{The Mass Functions of Blue and Red Galaxies}

\begin{figure}
\plotone{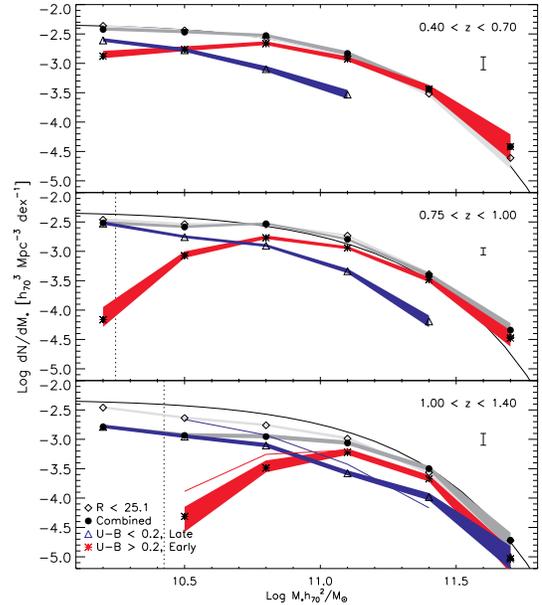}
\caption{Mass functions in three redshift bins partitioned by rest-frame
  $(U-B)$ color as described in the text.  Shading indicates the width
  of 1$\sigma$ error bars. The mass function for all galaxies in the
  primary spectroscopic sample is designated by solid circles. That for
  the deeper, $R_{AB} \leq 25.1$ sample, supplemented by photo-z's, is
  indicated by open diamonds.  Vertical dotted lines show estimates for
  mass incompleteness resulting from the \kband~magnitude limit.  In the
  high-redshift bin (bottom panel), thin red and blue lines trace the
  expected increase in the red and blue populations if the spectroscopic
  limit of $R_{AB} = 24.1$ were extended by one magnitude.  The isolated
  error bar in the upper right-hand portion of each plot indicates the
  estimated systematic uncertainty due to cosmic variance.  The solid
  curve in each panel shows the Schechter fit to the photo-$z$
  supplemented mass function in the first redshift interval.
\label{mfn_tot_col}}
\end{figure}

We now partition the total mass functions presented in Figure
\ref{mfn_tot_ref} into active and quiescent populations according to the
bimodality observed in the restframe $(U-B)$ color.  The results are
illustrated in Figure \ref{mfn_tot_col}, which for reference also plots
the total mass functions from Figure \ref{mfn_tot_ref}.  As before, the
vertical dotted lines indicate the onset of incompleteness resulting
from the \kband\ limit, and the solid curve is a Schechter fit to the
photo-$z$ supplemented total mass function in the first redshift
interval.

\begin{figure}
\plotone{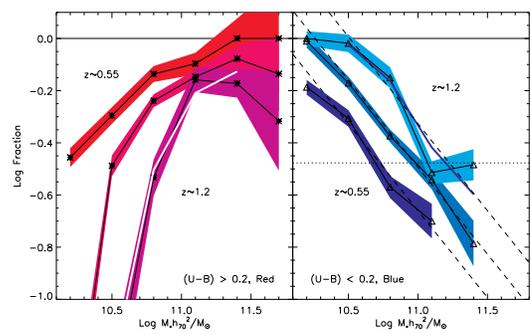}
\caption{Log fractional contribution of the red and blue populations to
  the total stellar mass function at various redshifts.  The relative
  contribution of red galaxies (left plot) increases with cosmic time,
  while that for blue galaxies decreases (right plot).  The expected
  high-$z$ fraction of red and blue types after correcting for $R$-band
  completeness (as in Figure \ref{mfn_tot_col}) is shown by a thin white
  line for the red population and a thin blue line for the blue
  population.  The corrected fractions from the $R_{AB} \leq 25.1$
  sample are entirely consistent with those observed in the primary
  $R_{AB} \leq 24.1$ spectroscopic sample.  Dashed lines in the
  right-hand panel show approximate fits to the declining fraction of
  blue galaxies. $M_Q$ is defined by where these lines cross the dotted
  horizontal line, which marks a fractional abundance of 1/3 the total
  value.  (see $\S$\ref{MQ}).
\label{frac_col}}
\end{figure}

As discussed above, the $R \leq 24.1$ limit introduces incompleteness in
the high redshift bin.  To mitigate this effect we derive a
color-dependent completeness correction for the high-$z$ bin based on
the photo-$z$ supplemented $R \leq 25.1$ sample.  Inferring the
restframe $(U-B)$ color for this sample is difficult because of
photometric redshift uncertainties.  Instead, we adopt the simpler
approach of applying a color cut in observed $(R-K)$, which, as shown in
Figure \ref{RK}, maps well onto the restframe $(U-B)$ color for the
high-$z$ spectroscopic sample.  We tune the $(R-K)$ cut so that the
resulting color-dependent mass functions of the $R_{AB} \leq 25.1$
sample match the spectroscopic ($R_{AB} \leq 24.1$) mass functions above
$M_* =10^{11.3}$\msun~where the spectroscopic sample is complete.  This
yields a value of $(R-K_s)_{2''} = 3.37$, consistent with the $(U-B)$
bimodality apparent in the color-magnitude diagram shown in Figure
\ref{UB}.

Using this observed $(R-K)$ color cut for the high-$z$ bin only, we  
show the color-dependent mass functions for the $R_{AB} \leq 25.1$ sample 
in Figure \ref{mfn_tot_col} as thin white and blue lines.  While these curves suffer
from their own uncertainties, such as photo-$z$ errors and a
less precise measure of color, they are useful for illustrating the
nature of the $R$-band incompleteness in the spectroscopic sample.  It
is important to note that while the deeper sample yields higher mass
functions for both the red and blue populations below $M_* \approx
10^{11.1}$\msun, the relative contribution of each one to the total mass
function is similar to what is observed in the spectroscopic sample.
This is shown more clearly in Figure \ref{frac_col}.  It should also be
noted that the $R_{AB} \leq 25.1$ sample itself is not complete,  
although the decreasing density of points near the $R_{AB} = 25.1$ limit in
Figure \ref{mass_complete} suggests it is largely complete over the  
range of stellar masses probed.

Figure \ref{mfn_tot_col} reveals several striking patterns.  A clear
trend is observed in which the abundance of massive blue galaxies
declines substantially with cosmic time, with the remaining bulk of the
actively star-forming population shifting to lower mass galaxies.  As
the abundance of the blue population declines, red galaxies, which
dominate the highest masses at all redshifts, become increasingly
prevalent at lower masses.  The two populations seem therefore to
exchange members so that the total number density of galaxies at a given
stellar mass remains fixed.  We also note the clear downward evolution
of the cross-over or transitional mass, $M_{tr}$, where the mass
functions of the two color populations intersect.  Above $M_{tr}$, the
mass function is composed of primarily red galaxies and below it blue
galaxies dominate.  We return to this behavior in $\S$\ref{MQ}.

Figure \ref{frac_col} shows these results in a different way.  Here, the
log fractional contribution from the red and blue populations are plotted in
the same panel so that the redshift evolution is clearer.  The
completeness-corrected color-dependent mass functions (with $R \leq 25.1$)
are shown as the solid white and blue lines.  Their overlap with the
shaded curves from the high-$z$ spectroscopic sample is remarkable and
indicates that unlike absolute quantities, relative comparisons between
mass functions drawn from the spectroscopic sample are not strongly
biased by the $R$-band mass completeness limit.  As noted previously,
plotting the relative fraction also removes the first order systematic
uncertainty from cosmic variance, making comparisons across the redshift
range more reliable.

The downsizing evolution in Figure \ref{mfn_tot_col} is now more
clearly apparent in Figure \ref{frac_col}.  The relative abundance of red
galaxies with $M_* \approx 6 \times 10^{10}$\msun~increases by a  
factor of $\approx$3 from $z \sim 1.2$ to $z \sim 0.55$.  At the same time, the
abundance of blue, late-type galaxies, which are thought to have
experienced recent star formation, declines significantly.

\subsection{Downsizing in Populations Defined by SFR and Morphology} 
\label{multi}

\begin{figure*}
\plotone{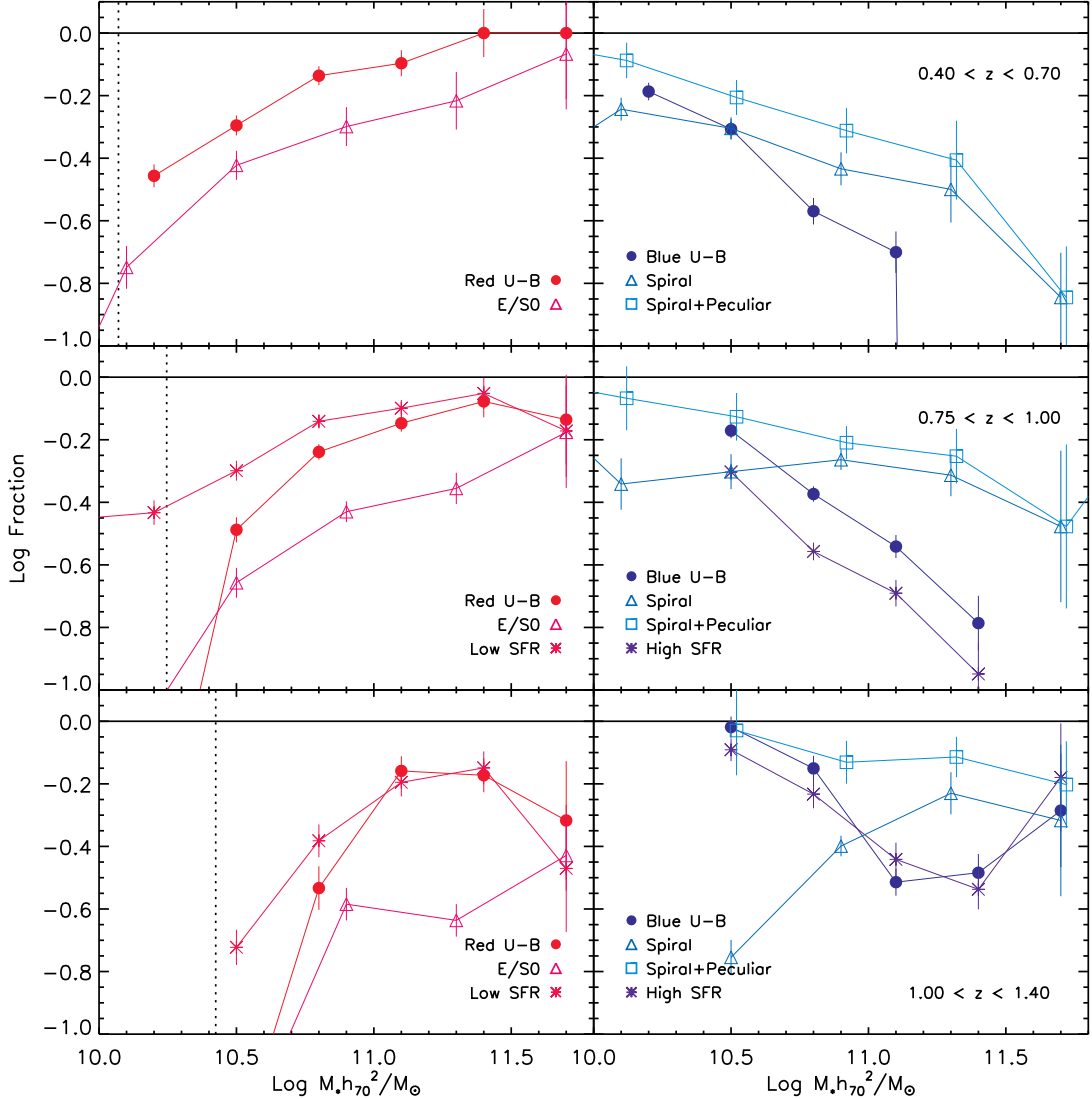}
\caption{Fractional contribution in log units to the total mass function
  from quiescent (left-hand plots) and active populations (right-hand
  plots) defined by rest-frame $(U-B)$ color, [OII] star formation rate,
  and morphology (see text for details).  Star formation rates based on
  [OII] are only available for galaxies with $z > 0.75$.  The
  morphological mass functions are those of \citet{bundy05}.  The
  \kband~completeness limits are indicated as in Figure
  \ref{mfn_tot_col}.\label{multi_plot}}
\end{figure*}

The patterns in Figure \ref{mfn_tot_col} and \ref{frac_col}
are also apparent when the galaxy population is partitioned by other
indicators of star formation.  This is demonstrated in Figure
\ref{multi_plot} which shows the fractional contribution (in log
units) of active and quiescent  populations 
to the total mass function. The ``blue'' and ``red'' samples defined 
by restframe $(U-B)$ color and shown in Figure \ref{mfn_tot_col} 
are reproduced here and indicated by solid circles.

For the two redshift intervals with $z > 0.75$, we have plotted
contributions from samples with high and low [OII]-derived SFRs.  We
divide this sample at 0.2 \msun~yr$^{-1}$, which is the median SFR of
the star-forming population at $0.75 < z < 1.0$.  This imposes a more
stringent criterion than the restframe $(U-B)$ cut, which counts
galaxies with only moderate or even recent star formation as
``late-type.''  Not surprisingly, the middle redshift bin contains fewer
high-SFR galaxies compared to blue $(U-B)$ systems and more low-SFR
galaxies than red systems, although the mass-dependence observed with
either criteria is qualitatively similar.

In the high-$z$ bin, the populations defined by color and [OII] track each
other more closely.  Not only does this confirm that the mass-dependent
evolution seen in Figure \ref{mfn_tot_col} is reproduced when the sample
is divided by the [OII] SFR, but it also indicates that the average star
formation rate is {\em higher} in this redshift bin.  More of the  
``blue'' population is now above the SFR cut as compared to the middle redshift
bin.  This evolution in the observed SFR will be discussed in detail in
Noeske et al. (in preparation).

It is helpful also to understand how these trends relate to earlier work
motivated by understanding the role of morphology in downsizing.  Figure
\ref{multi_plot} also plots the contribution from
morphologically-defined populations, drawing from the sample of
\citet{bundy05} which has been adjusted to the $h=0.7$ cosmology used
here (we note that the recent addition of HST/ACS imaging in the EGS provides an
opportunity to extend this morphological comparison in the future).  In
\citet{bundy05}, morphologies were determined visually using HST/ACS
imaging data from the GOODS fields \citep{giavalisco04} and were divided
into three broad classes: E/S0, spirals, and peculiars.  The fractional
contribution from the spiral and spiral+peculiar samples are plotted in
Figure \ref{multi_plot} for comparison to the late-type populations
described above.  The E/S0 fraction is compared to the early-types.  It
should be noted that the smaller sample size of the \citet{bundy05} data
leads to greater uncertainties and larger effects from cosmic variance.

With these caveats, there is quite good agreement in the mass-dependent
evolutionary trends between the morphological and color/SFR selected samples.
In detail, the fraction of ellipticals is systematically lower than the
red/low-SFR populations while the fraction of spirals+peculiars is
systematically higher than the blue/high-SFR galaxies.  This suggests
that the process that quenches star formation and transforms late-types
into early-types operates on a longer timescale for morphology than it
does for color or SFR.  We return to this point  in $\S$\ref{discussion}.
We also note that spiral galaxies do not always exhibit star formation
and can be reddened by dust while some ellipticals have experienced
recent star formation \citep[e.g.,][]{treu05:downsize} that could lead to
bluer colors.  

\subsection{Quantifying Downsizing: the Quenching Mass 
Threshold, $M_Q$}\label{MQ}

\begin{figure}
\plotone{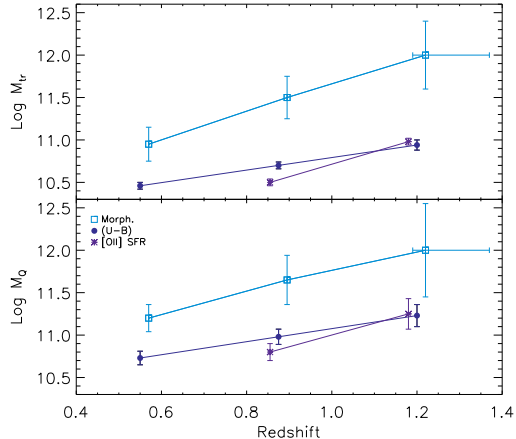}
\caption{Redshift evolution of the characteristic transitional mass,
  $M_{tr}$ (top plot), as well as the quenching mass, $M_Q$ (bottom
  plot).  The plotted redshift values indicate the approximate average
  redshift in each bin.  In both plots, the observed behavior is shown for
  the population partitioned according to morphology (light, open
  squares), restframe $(U-B)$ color (solid circles), and [OII] SFR
  (asterisks).
\label{Mtr_MQ}}
\end{figure}

Several authors have identified a characteristic transition mass,
$M_{tr}$, which divides the galaxy stellar mass function into a
high-mass regime in which early-type, quiescent galaxies are dominant
and a low-mass regime in which late-type, active galaxies are dominant
\citep[e.g.,][]{kauffmann03:bimodal, baldry04, bundy05}.  Using our
various criteria, the downward evolution of this transitional mass with
time is clearly demonstrated in the upper panel of Figure \ref{Mtr_MQ}.
For the color based mass functions, good agreement is found in
comparison to the COMBO17 results discussed in \citep{borch06}.  The
morphological sample is taken from \citet{bundy05}, where we have
grouped spirals and peculiars into one star-forming population and
compared its evolution to E/S0s.  In the high-$z$ bin, $M_{tr}$ for the
morphological sample occurs beyond the probed stellar mass range and so
we have extrapolated to higher masses to estimate its value (this
uncertainty is reflected in the horizontal error bar at this data
point).

The color-defined $M_{tr}$ shows a  redshift dependence of
$M_{tr} \propto (1+z)^{4}$, similar to that for the morphological
sample. Stronger evolution is seen for the the [OII]-defined samples
as expected if evolution is more rapid for the most active sources.
As discussed in $\S$\ref{multi}, the mass scale of morphological evolution 
is approximately 3 times larger ($\approx$0.5 dex) than that defined by 
color or [OII].  We also note from Figure \ref{mfn_tot_col} that $M_{tr}$ does 
not change appreciably when the $R$-band mass incompleteness is 
corrected in the high-$z$ bin.

While the evolution in $M_{tr}$ is illustrative of downsizing, since its
definition in terms of physical processes is completely arbitrary
(equality in the relative mass contributions of two populations), its
significance is not clear.  We prefer to seek a quantity that clearly
describes the physical evolution taking place. Accordingly, we introduce
and define a {\em quenching mass limit}, $M_Q$, as that mass above which
star formation is suppressed in galaxies.  This threshold is a direct
byproduct of the mechanism that drives downsizing.  We define $M_Q$ by
noting the mass at which a line fit to the declining fraction of
star-forming galaxies (Figure \ref{frac_col}) drops below 1/3.  We apply
this simple definition to the other measures of star-forming populations
and plot the resulting values of $M_Q$ in the bottom panel of Figure
\ref{Mtr_MQ} and list them in Table \ref{MQ_table}.  The relative
behavior of differently classified populations is similar to the top
panel, but the physical mass scale associated with $M_Q$ is a factor of
2--3 higher than that of $M_{tr}$.  We find an approximate redshift
dependence of $M_Q \propto (z+1)^{4.5}$, similar to the dependence of
$M_{tr}$.

$M_Q$ is a useful quantity because it reveals the masses at which
quenching operates effectively as a function of redshift.  What is more,
blue and star-forming galaxies at the quenching mass also tend to be
among the reddest and least star-forming as compared to the full
late-type population as a whole at each redshift, suggesting such
galaxies are in the process of becoming early-types.  As we discuss
below, the quantitative evolution of $M_Q$ strongly constrains what
mechanism or mechanisms quench star formation and also provides a
convenient metric for testing galaxy formation models.

\begin{deluxetable}{lccc}
\tablecaption{The Quenching Mass Threshold, $M_Q$}

\tablewidth{0pt}
\tablecolumns{4}
\tablehead{
\colhead{Redshift} & \colhead{Blue $(U-B)$} & \colhead{High SFR} &
\colhead{Spirals+Peculiars} }

\startdata
$0.4 < z < 0.7$  & $10.73 \pm 0.08$ & ---              & $11.20 \pm
0.16$ \\
$0.75 < z < 1.0$ & $10.98 \pm 0.09$ & $10.80 \pm 0.10$ & $11.65 \pm
0.29$ \\
$1.0 < z < 1.4$  & $11.23 \pm 0.13$ & $11.25 \pm 0.18$ & $12.00 \pm
0.55$ \\
\enddata
\label{MQ_table}
\tablecomments{$M_Q$ is given in units of $\log (M
   h_{70}^2/M_{\odot})$.  Morphology data comes from \citet{bundy05}.}
\end{deluxetable}

\subsection{The Environmental Dependence of Downsizing}

\begin{figure*}
\plotone{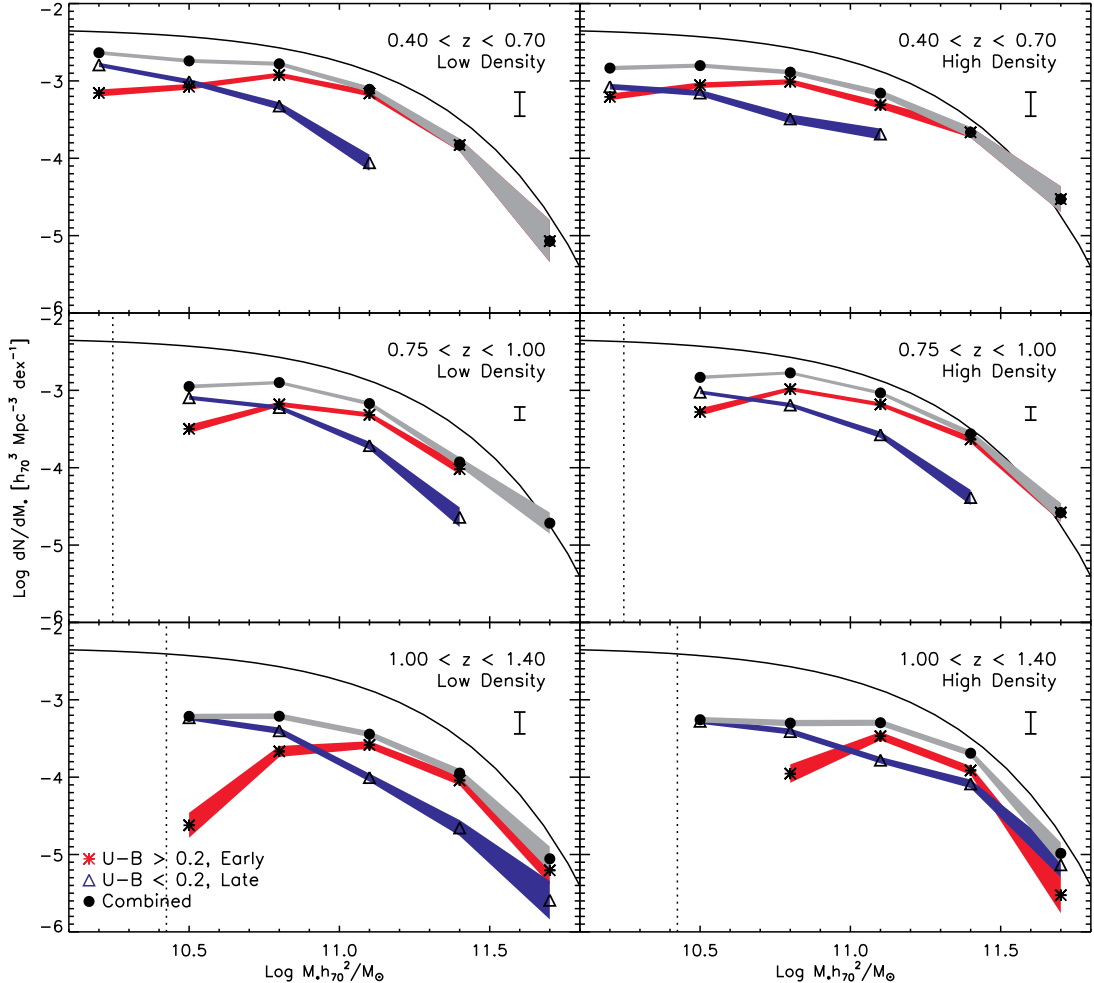}
\caption{Stellar mass functions in the three redshift bins at above- and
  below-average density partitioned by rest-frame $(U-B)$ color.  The
  vertical dotted lines show the \kband~magnitude completeness limits.
  The estimated cosmic variance is designated by the isolated error bar.
  Significant incompleteness from the $R$-band limit is expected in the
  high redshift bin, but it is not possible to apply corrections as in
  Figure \ref{mfn_tot_col} because the local density can only be
  measured in the spectroscopic sample. The solid line shows the
  Schechter fit to the photo-$z$ supplemented total mass function in our
  lowest redshift bin.\label{mfn_col_dens}}
\end{figure*}

We have so far considered the mass-dependent evolution of late- and
early-type populations integrated over the full range of environments
probed by the DEEP2 Redshift Survey.  We now divide the sample by
environmental density to investigate how this evolution depends on
the local environment.

In Figure \ref{mfn_col_dens}, we plot galaxy stellar mass functions for
the samples shown in Figure \ref{mfn_tot_col}, split into low-density
environments on the left-hand side and high-density environments on the
right-hand side.  The density discriminant is simply the median density
(indicated by an overdensity of zero in Figure \ref{dens_dist}), so all
galaxies are included.  The evolution of the total mass functions in the
two density regimes follows similar patterns as for the combined sample
in Figure \ref{mfn_tot_col}.  Furthermore, the relative contribution of
active and quiescent galaxies, as quantified by $M_{tr}$ and $M_Q$,
exhibit the same kind of downsizing signal seen in Figure
\ref{mfn_tot_col}.  Crucially, there is little difference (less than
25\%) in the value of $M_{tr}$ between the below- and above-average
density samples.

\begin{figure}
\plotone{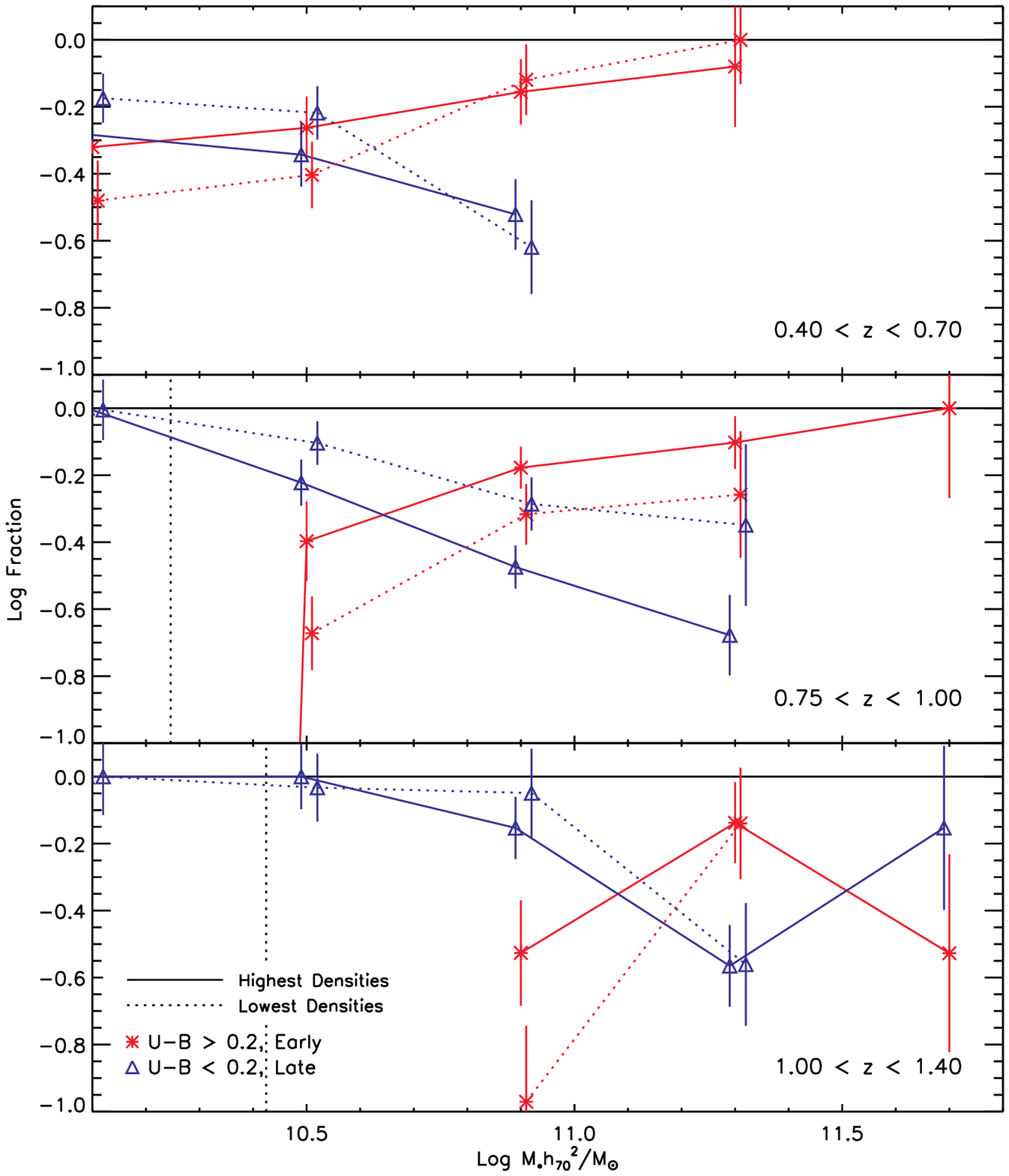}
\caption{Fractional contribution in log units to the total mass
   function.  In each of three redshift bins solid lines connect points
   representing the extreme high-end of the density distribution,  
   while dotted lines indicate the extreme low-end.  The \kband~completeness
   limits are indicated as vertical dotted lines.\label{frac_densX}}
\end{figure}

The absence of a strong environmental trend in the downsizing pattern of
massive galaxies is a surprising result that warrants further scrutiny.
Although no strong dependence was seen in the Fundamental Plane analysis
of \citet{treu05:downsize}, their density estimates were much coarser
and based on photometric redshifts.  Local work utilizing the SDSS by
\citet{kauffmann04} demonstrated differences in galaxy properties as a
function of environment, but covered a much larger dynamic range in
density than is accessible with our sample.  We emphasize that the
results of this work focus primarily on field galaxies, and, with the
typical volumes probed ($\sim$10$^6$ Mpc$^3$), only rich groups are
included at the highest densities.  DEEP2 is too small to probe cluster
scales.

It is important to note that individual density measurements, even in
our spectroscopic analysis, can exhibit uncertainties of a factor of 3,
so there is some overlap expected between the above- and below-average
density regimes in Figure \ref{mfn_tot_col}.  However, because the total
mass functions exhibit different abundances---with a factor of $\sim$2
more high-mass galaxies in the above-average density regime---the above-
and below-average density subsamples successfully probe different populations.
The higher abundance of massive galaxies at increasing densities agrees
well with trends in local studies \citep[e.g.,][]{balogh01, balogh04,
  baldry04, kauffmann04}.

Still, the comparison above is inevitably dominated by galaxies near the
median density (Figure \ref{dens_dist}), and so we adopt a second, more
stringent approach that divides the distribution into three, more
extreme density bins. In this way, we can sample the full dynamic range
of our survey.  As described in $\S$\ref{density}, we define a second
set of density thresholds at $\pm$0.5 dex from the median density and
construct mass functions for the extreme high- and low-density regimes.
We plot the results in Figure \ref{frac_densX}, which again shows the
fractional contribution of the red and blue populations to the total
mass function.  This time, the samples are also divided by density, with
the high density points connected by solid lines and the low density
points connected by dotted lines.

Figure \ref{frac_densX} illustrates a more substantial environmental
effect in the mass-dependent evolution of the sample that in each
redshift interval is suggestive of local environmental trends
\citep[e.g.,][]{kauffmann04}.  Our observations allow us to investigate
how these trends evolve with time.  We find that the rise of the
quiescent population and the evolution of $M_{tr}$ appears to be
accelerated in regions of high environmental density.  This effect does
not depend on the particular choice of the density threshold, although
the differences between the two environments grow as more of the sample
near the median density is excluded from the analysis.  The difference
in $M_{tr}$ between these two environments is roughly a factor of 2--3.
We caution that interpreting the results in the high-$z$ bin is
difficult because the effects of completeness and weighting are most
important here.  However, the fact that the $M_{tr}$ values in the two
density regimes are more similar in the high-$z$ bin (an offset of
$\sim$0.1 dex) might suggest that the structural development or physical
processes that lead to the density dependence at lower redshifts are not
fully in place before $z \sim 1$.

The environmental dependence observed in Figure \ref{frac_densX} does
not occur because of the high-density regime alone, as one might expect
given the potential presence of dense structures in the sample.  The
environmental effect is less strong but still apparent in comparisons
between the high and middle density regimes as well as between the
middle and low density regimes.

In summary, our various measures of downsizing, $M_{tr}$ and $M_Q$,
depend strongly on redshift (they evolve by factors of $\sim$3--5 across
the full redshift range), but less so on environmental density for most
field galaxies (less than 25\%).  In support of previous inferences on
the role of environment based on local studies \citep[e.g.,][]{balogh04,
  kauffmann04, blanton04}, this direct measurement of density-dependent
evolution is particularly striking and serves to emphasize that galaxy
mass, not environmental location, is the primary parameter governing the
suppression of star formation and hence producing the signature of
downsizing.

\section{Discussion}\label{discussion}

\subsection{The Rise of Massive Quiescent Galaxies}

Figures \ref{mfn_tot_col} and \ref{frac_col} demonstrate a clear
feature of the downsizing signal observed since $z \sim 1$, namely the
increase in the number density of massive quiescent galaxies.  Although
our results are consistent with previous studies which found a rise in
the red galaxy abundance of a factor of $\approx$2--6, depending on mass
\citep{bell04:redevol}, the present work represents a significant step
forward not only in its statistical significance and precision by virtue
of access to the large spectroscopic and infrared data set, but also in
clearly defining the mass-dependent trends.

In discussing our results we begin by considering the processes that
might explain the present-day population of early-type galaxies.  In
order to reconcile the significant ages of their stellar populations
implied by precise Fundamental Plane studies
\citep[e.g.,][]{treu05:downsize, vdwel05} with hierarchical models of
structure formation, \citet{bell05:dry}, \citet{faber05},
\citet{vandokkum05:merge} and others have introduced the interesting
possibility of ``dry mergers''---assembly preferentially progressing via
mergers of quiescent sub-units.

While dry mergers clearly occur \citep{tran05, vandokkum05:merge}, our
results suggest that they cannot be a substantial ingredient in the
assembly history of massive quiescent galaxies.  As shown in Figure
\ref{mfn_tot_col}, the observed increase in the number of quiescent
systems is almost perfectly mirrored by a decline in star-forming
galaxies such that the total mass function exhibits little evolution
over $0.4 < z < 1.4$ compared to that of the red and blue populations.
A simple transformation of one into the other is sufficient to high
precision.  For example, simply interchanging the numbers of red and
blue galaxies in the high-$z$ bin in the mass range $10^{11.2} <
\log{M_*/M_{\odot}} < 10^{11.8}$, leads to a prediction for the number
density in the middle-$z$ bin that is accurate to within $\sim$25\%,
well within the cosmic variance uncertainty.  In addition, galaxies
defined as blue or late-type with masses near $M_Q$ tend to be among the
reddest and least star-forming, suggesting they are likely candidates
for becoming early-type systems.

It is conceivable that the dry merger rate is mass-dependent and 
conspires to move galaxies along the mass function 
in a way that leaves its shape preserved.  This would imply the presence 
of massive galaxies at low redshift that are not seen in our sample.  
For example, using the approximate dry merger rate estimate from 
\citet{bell05:dry} of 1.3 mergers every 6.3 Gyr,  $\sim$25\% of red
galaxies would have to experience a major dry merger (in reality, 
this rate, calculated over $0 < z < 0.7$, might be expected to be 
higher at $z \sim 1$) between the high-$z$ and middle-$z$ bins.  
If we apply this rate of assembly to the red population in our high-$z$ 
bin at $\log (M_*/M_{\odot}) = 11.4$, we find that the total mass 
function in the middle-$z$ bin at $\log (M_*/M_{\odot}) = 11.7$ 
should be higher by approximately 60\% compared to what is 
observed.  While this one data point represents only a 
$\approx$2$\sigma$ result, similar arguments apply across 
the mass function and between the middle-$z$ and low-$z$
bins.  More detailed modeling of the effect of merger
rates on the mass function will be presented in a further paper 
(Bundy et al., in preparation).

Most difficult for the dry merger hypothesis as a key feature of galaxy
formation is the weak environmental dependence we observe in the
downsizing signal. Given a mechanism has to be found to preferentially
bring quiescent sub-units together, one would expect a strong density
dependence in the dry merger rate. By contrast, in our large sample it
is clear that the majority of quiescent galaxies were assembled in a
manner more sensitive to mass than environment.

Finally, our results show that downsizing is not only a feature of the
star formation histories of massive quiescent galaxies
\citep[e.g.,][]{treu05:downsize, vdwel05}, but is also apparent in the
way their abundance increases.  This observation is of particular
interest for galaxy formation models based on the hierarchical
$\Lambda$CDM framework.  By incorporating AGN feedback effects into the
semi-analytic models from the Munich group, \citet{delucia05} predict
star formation histories for massive ellipticals that follow the
downsizing trend in which more massive galaxies host older stellar
populations.  However, as shown in Figure 5 from \citet{delucia05},
these models still predict a hierarchical mass assembly history for
ellipticals.  The problem in comparing the predictions of
\citet{delucia05} to our results (Figure \ref{frac_col}, for example) is
that the simulated ellipticals in \citet{delucia05} are defined by their
appearance at $z=0$, not the redshift at which we observe them.  While
the observations presented here suggest a top-down pattern in the
growing abundance of early-types, it is not yet clear how significant a
problem this may be for current semi-analytic models.

\subsection{The Origin of Downsizing}\label{origin}

\begin{figure}
\plotone{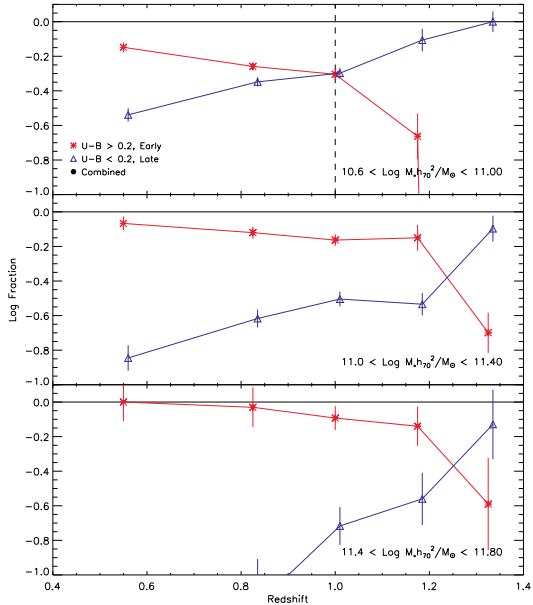}
\caption{Fractional contribution in log units to the total mass
  function, divided into stellar mass bins and plotted as a function of
  redshift.  The plotted redshift values indicate the approximate
  average redshift in each bin.  The vertical dashed line in the top
  panel indicates the onset of incompleteness from the $R$-band limit
  which occurs above the redshift indicated.  The \kband\ completeness
  limit does not affect the results plotted in this
  figure.\label{mfn_evol_col}}
\end{figure}

The results of this paper reveal important clues as to the nature of downsizing
and, via a clear measurement of the trends, will assist in constraining and  
ruling out several of the popular explanations.  A detailed comparison
with such models is beyond the scope of the current paper but we discuss
some of the key issues here.

First we wish to dismiss a possible suggestion that our discovery of the
quenching mass threshold is somehow an artifact of our selection
process.  For example, it might be argued that our result could arise
from a uniform decrease in the incidence of star formation at all masses
combined with a survey selection effect in which rare, massive objects
are seen only at higher redshifts because of the larger volumes probed.
Our results exclude this possibility.  First, our sampled volumes are
relatively similar (only a factor of $\sim$4 difference between low-$z$
and high-$z$), and we demonstrate that the fraction of star-forming
galaxies depends not only on redshift but mass as well.  Figure
\ref{mfn_evol_col} illustrates this point.  Here we divide the sample
into smaller redshift intervals and follow the evolution of the red and
blue populations in the three largest stellar mass bins, charting the
fractional contribution of the two populations to the combined mass
function.  The highest mass bin contains the largest fraction of
quiescent galaxies at almost every redshift, and the transformation of
the active population into passively-evolving systems occurs first in
the high mass bin and later at lower masses.  The rate of change in the
incidence of star formation is clearly mass-dependent.

Turning now to physical explanations, we have shown in $\S$\ref{MQ} that
the quenching mass threshold, parametrized by $M_Q$, provides a very
useful description of how the fraction of star-forming galaxies evolves.
The question then is what mechanism is responsible for this quenching?
Can it adequately reproduce the quantitative trends observed, for
example the environmental dependence? Merging may provide a starting
point for answering this question and explaining the transformation of
late-types into early-types.  Merging between disk systems has long been
thought to be an important mechanism by which ellipticals form
\citep[e.g.,][]{toomre77, barnes91, springel05:feedback} and the similar
behavior of the morphological and color-defined values of $M_Q$ in
Figure \ref{Mtr_MQ} suggests that the same process governing the growth
of ellipticals may also broadly explain the rise of quiescent galaxies
in general.  However, as argued above, significant merging is likely to
affect the {\em shape} of the total mass function which does not appear
to evolve strongly in our sample. Merging may be occurring at masses
below our completeness limit, but the observed evolution in the relative
mix of early- and late-type galaxies suggests a process that quenches
star formation first with morphological transformation following later.
This would lead to higher values for $M_Q$ in the morphological samples,
as observed.

Merger-triggered quenching has further difficulties. Fundamentally, the
hierarchical merging of dark matter halos is expected to proceed from
low mass to high mass, not the other way around.  One solution to this
difficulty would be to appeal to the fact that merging and assembly
rates are accelerated in regions of high density
\citep[e.g.,][]{delucia04} which also host the most massive
systems. Over a range of environments, downsizing could arise naturally
from the fact that massive galaxies live in these accelerated
environments.  However, we find no significant density dependence in the
bulk of our sample (Figure \ref{mfn_col_dens}), with downsizing
appearing in all environments.  This suggests that density-dependent
merger rates are not the answer and that an internal feedback process on
galactic scales is largely responsible for driving the downsizing
pattern.

Many groups have recently suggested that internal AGN feedback may be
the missing ingredient. Triggered perhaps by merging, energy deposited
by the AGN effectively quenches further star formation, eventually
transforming late-type galaxies into early-types
\citep[e.g.,][]{silk98}.  There are at least two current implementations
of this process.  In the explosive model, energy from an intense
``quasar phase'' (usually triggered by merging) heats and expels gas in
the halo that would otherwise be available for star formation
\citep{granato04, springel05:ellip, scannapieco05}.  While inspired by
hydrodynamic simulations, it is not clear how the downsizing mass scale
could arise from this model or what prevents future cooling of the
expelled gas.  In the ``radio feedback'' model, low luminosity AGN
energy couples to hot gas in halos with masses greater than $M_{shock}$
in which the cooling time is longer than the dynamical time
\citep[see][]{dekel04}.  This results in a long-term suppression of star
formation in massive halos and, on even larger scales, cooling flows
\citep[e.g.,][]{croton05, delucia05, bower05, cattaneo06}.  These models
produce older stellar populations in more massive galaxies but still
predict that low mass early-types (as defined at $z=0$) assemble first
(see Figure \ref{frac_col}).  While a possible mass scale is provided in
$M_{shock}$ that may help drive the downsizing signal, the physical
mechanism that couples the radio feedback to the hot gas is not well
understand.  Beyond simulations and formation models incorporating AGN,
additional evidence for the role of AGN feedback comes from
phenomenological studies that relate observations of black hole mass,
the quasar luminosity function, and the properties of galaxies in a
consistent framework \citep[e.g.,][]{merloni04, hopkins05:red,
  merloni06, hopkins06:scales}.

Regardless of the physical explanation for downsizing in the star
formation activity of galaxies (and several may be necessary), it is
clear that precise quantitative measures of the evolving mass
distribution and its dependence on the basic parameters explored here
will provide the ultimate test of these theories.

\subsection{Reconciling Downsizing with the Hierarchical Structure Formation}

In this section we step back from specific models of galaxy formation
and explore in a more general sense the kinds of processes that may
reconcile downsizing with hierarchical structure formation.  Over the
past decade there has been strong confirmation from many independent
observations that the large-scale structure of the universe matches the
basic predictions of the cold dark matter (CDM) model in which dark
matter ``halos'' harbor galaxies and grow through constant, hierarchical
merging with other halos.  The masses of all halos increase
monotonically with time, but at every epoch, it is the largest halos
that are growing most actively and that are also the last to have
assembled.  Thus at first examination, downsizing, in which star
formation proceeds from high-mass to low-mass systems, seem at odds with
the CDM picture.

Several processes may be contributing to reversing the bottom-up trend
in CDM structure formation to produce what appears to be a top-down
pattern in galaxy formation. The first is a gradual effect resulting
from the accelerating expansion of the universe which causes halo growth
rates to slow once the universe reaches a scale factor $(1 + z)^{-1} >
\Omega_m$. The second is the physics of gas cooling, which selects a
characteristic mass scale for galaxy formation \citep{rees&ostriker77,
  silk77, white&rees78}. Gas cannot cool rapidly, and by implication
stars cannot form efficiently, until structure formation produces halos
with virial temperatures in excess of $10^4$~K . This sets the epoch for
the onset of galaxy formation at $z\sim15$--20. Once gas temperatures
increase to $10^6$--$10^7$~K, cooling once again becomes inefficient,
turning off star formation in the most massive halos. This then marks
the end of galaxy formation, as more and more mass builds up in group
and cluster halos above this cooling limit.

More detailed numerical or semi-analytic models of galaxy formation show
that the cooling delay alone is insufficient to reduce star formation to
observed levels, particularly in massive halos, and that other forms of
feedback are required.  While the details remain controversial
\citep[e.g.,][]{benson03}, the net effect of this feedback is to place an
upper limit on the range of halo masses over which active star formation
can take place. This limit helps explain why star formation in galaxies
is rarer at the present-day than it was at $z\sim$1--2.  However, as
discussed in $\S$\ref{origin}, it is less obvious how to explain the
observed decline in the {\em mass scale} of star-forming galaxies.

To help gain insight into this question, we can attempt to relate
various galaxies in our sample to dark matter halos. Models of the halo
occupation distribution function predict that galaxies with the range of
stellar masses sampled here ($\log (M/M_{\odot})\sim 10$--12,
corresponding to $\log L_{b_J}\sim10$--11.2) should reside in dark
matter halos of mass $10^{12}$--$10^{15} M_{\odot}$
\citep[e.g.,][]{yang03, cooray05}, and furthermore that 75-80\% of these
galaxies will be ``central,'' that is the dominant galaxies within their
halo, rather than satellites \citep{cooray05:anatomy}. These models
suggest that the objects in the three mass bins in Figure
\ref{mfn_evol_col} correspond approximately to central galaxies in
galaxy, group, and cluster halos.

\begin{figure}
\plotone{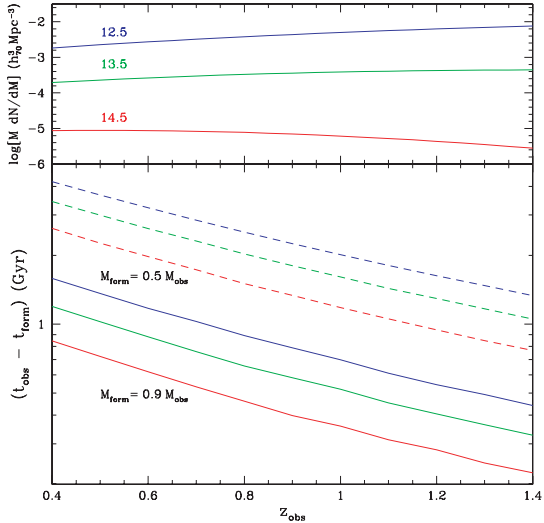}
\caption{(Upper panel) The abundance of halos likely to host central galaxies 
in the three mass ranges plotted in Figure \ref{mfn_evol_col} versus redshift. 
The curves are labeled with $\log(M/M_{\odot})$. (Lower panel) Mean age of 
these systems as a function of their observed redshift. The mean age is defined 
as the time elapsed since half of the systems had built up a fraction $f$ of the 
mass they have at $z_{obs}$. Solid curves show the age for $f=0.9$ and dashed 
curves that for $f=0.5$. \label{mean_age}}
\end{figure}

In the top panel of Figure \ref{mean_age} we show the comoving number
density of halos of mass $\log (M/M_{\odot})\,=\,12.5, 13.5$ and 14.5 as
a function of redshift (three lines, from top to bottom).  The numbers
are roughly consistent with the comoving number densities of galaxies in
our three mass bins, although the most massive stellar objects are more
abundant than $10^{14.5} M_{\odot}$ halos and may therefore reside in
slightly less massive systems.  The bottom panel shows the mean ages of
halos in the three mass bins as a function of observed redshift. The
mean age here is defined as the time elapsed since half of the halos in
that mass range had first built up 90\% or 50\% of their current mass in
a single progenitor (solid and dashed curves respectively), calculated
using equation 2.26 from \citet{lacey&cole93}.

By comparing Figure \ref{mean_age} to Figure \ref{mfn_evol_col} we can
gain insight into the relationship between the assembly of dark matter
halos and the process that drives the observed mass-dependent decline of
star-forming galaxies.  The fraction of the most massive blue galaxies
declines from nearly 100\% to $\sim$15\% by $z \approx 1$ in Figure
\ref{mfn_evol_col}.  For the least massive galaxies, this same decline lasts
over the full redshift range of our sample, implying a timescale 2.5
times longer.  Coupled with Figure \ref{mean_age}---which indicates that
the least massive halos in our sample are about twice the age of the
most massive halos---this suggests a process that, for the low-mass
systems, is at least 5 times slower than the growth of the dark matter
halos in which they reside.  Since global dynamical timescales should be
independent of halo mass at a given redshift, this suggests that the
quenching mechanism is strongly mass-dependent with the potential for
different physical processes acting in different mass ranges.

\section{Conclusions}\label{conclusions}

Using a large sample that combines spectroscopy from the DEEP2 Galaxy Redshift Survey
with panoramic near-IR imaging from Palomar Observatory, we have investigated the
mass-dependent evolution of field galaxies over $0.4 < z < 1.4$.  We
have constructed stellar mass functions for active  and quiescent populations,
defined in several ways, and divided into different samples according to
accurate measures of the environmental density determined from the extensive
spectroscopic data.  We summarize our conclusions below:

\begin{itemize}
\item The mass functions of active and quiescent galaxies integrated
   over all environments conclusively demonstrate a downsizing signal.  We  
   quantify this by charting the evolution in a ``quenching 
    mass,'' $M_{Q}$, which describes the mass scale above which  
   feedback processes suppress star formation in massive galaxies.  We find that
   $M_{Q} \propto (1+z)^{3.5}$ with a factor of $\approx$3 decrease
   across the redshift range probed.

\item The growth in the abundance of quiescent or ``early-type''
  galaxies occurs first at the highest masses and then proceeds to lower
  mass systems.  The relative abundance of early-types with $M_* \sim
  10^{10}$\msun~has increased by a factor of $\approx$3 from $z \sim
  1.2$ to $z \sim 0.55$, whereas the total mass function exhibits little
  evolution in shape and normalization (less than 0.2--0.3 dex).  This
  implies that early-type systems result largely via the transformation
  of active star-forming galaxies, indicating that ``dry mergers'' are not
  a major feature of their assembly history.

\item Alternative ways of dividing active and quiescent galaxies, including
  the use of [OII] equivalent widths and HST-derived morphologies, show qualitatively 
  similar mass-dependent evolution and quenching. Interestingly, we observe
  that morphological evolution appears to take place on longer timescales
  than changes in the apparent star formation rate which operate at
  lower mass scales at each redshift.

\item For the majority of galaxies in our sample, downsizing shows
  little dependence on environment.  An environmental signal is apparent
  when the extremes of the density distribution are compared.  In this
  case, downsizing in high-density regimes appears moderately
  accelerated compared to low-density ones, with values of $M_{tr}$
  lower by a factor of $\sim$2.

\item We discuss several possibilities for the origin of downsizing
   based on our results.  We clearly rule out a scenario in which the
   incidence of star formation decreases uniformly for galaxies at all
   masses.  The weak density dependence also argues against explanations that
   rely on the accelerated assembly of structure in dense environments,
   favoring internal mechanisms instead.

\item Through comparisons to the expected behavior of dark matter halos,
  we argue that the dynamical timescale resulting from the growth of
  structure is at least 5 times longer in galaxies hosted by halos with
  $\log M/M_{\odot}\approx 12.5$ ($\log M_*/M_{\odot} \approx 10.8$)
  compared to $\log M/M_{\odot}=14.5$ ($\log M_*/M_{\odot} \approx
  11.6$).  Because global dynamical scales are also independent of halo
  mass at a given redshift, this suggests that the quenching mechanism
  is strongly mass-dependent with the potential for different physical
  processes acting in different mass ranges.

\end{itemize}

There are two obvious avenues for further studies of downsizing.  In a
forthcoming paper (Bundy et al., in preparation) we discuss the
constraints on merging and the growth of galaxies determined by our
observations of the total mass function.  This will help dissect the
role of merging in driving downsizing.  In the near future, it will also
be possible to chart the incidence of AGN among the galaxy population
and compare it to the incidence of star formation to probe the link
between quenching and AGN.  The significant Chandra follow up
observations currently underway in the EGS will make that field
particularly exciting for such work.  Other efforts from the DEEP2
collaboration have provided new insight on environmental trends
\citep{cooper06} and will focus on precise measures of the evolving star
formation rate (Noeske et al., in preparation).

\acknowledgments

We are very grateful to the referee, Jarle Brinchmann, for very useful
comments and suggestions that have strengthened this work.  We also
thank Nick Kaiser for generously providing the optical photometry in
this paper.  The Palomar Survey was supported by NSF grant AST-0307859
and NASA STScI grant HST- AR-09920.01-A.  Support from National Science
Foundation grants 00-71198 to UCSC and AST~00-71048 to UCB is also
gratefully acknowledged.  A.L.C. is supported by NASA through Hubble
Fellowship grant HF-01182.01-A, awarded by the Space Telescope Science
Institute, which is operated by the Association of Universities for
Research in Astronomy, Inc., for NASA, under contract NAS 5-26555.  We
wish to recognize and acknowledge the highly significant cultural role
and reverence that the summit of Mauna Kea has always had within the
indigenous Hawaiian community. It is a privilege to be given the
opportunity to conduct observations from this mountain.



\end{document}